\documentclass[pra,twocolumn,showpacs,preprintnumbers,amsmath,amssymb,superscriptaddress]{revtex4}

\usepackage{color}
\usepackage{graphicx}% Include figure files
\usepackage{dcolumn}% Align table columns on decimal point
\usepackage{bm}% bold math
\usepackage{hyperref}
\usepackage{soul}
 \usepackage[T1]{fontenc}

\DeclareMathOperator{\sgn}{sgn}
\DeclareMathOperator{\Tr}{Tr}

\begin{document}

%\preprint{APS/123-QED}

\title{Two-axis two-spin squeezed states}

\author{Jonas Kitzinger}
\affiliation{State Key Laboratory of Precision Spectroscopy, School of Physical and Material Sciences, East China Normal University, Shanghai 200062, China}
\affiliation{New York University Shanghai, 1555 Century Ave, Pudong, Shanghai 200122, China} 
\affiliation{Humboldt-Universit\"at zu Berlin, Institut f\"ur Physik, Newtonstra{\ss}e 15, 12489 Berlin, Germany}

\author{Manish Chaudhary}
\affiliation{State Key Laboratory of Precision Spectroscopy, School of Physical and Material Sciences, East China Normal University, Shanghai 200062, China}
\affiliation{New York University Shanghai, 1555 Century Ave, Pudong, Shanghai 200122, China}

\author{Manikandan Kondappan}
\affiliation{State Key Laboratory of Precision Spectroscopy, School of Physical and Material Sciences, East China Normal University, Shanghai 200062, China}
\affiliation{New York University Shanghai, 1555 Century Ave, Pudong, Shanghai 200122, China} 

\author{Valentin Ivannikov}
\affiliation{State Key Laboratory of Precision Spectroscopy, School of Physical and Material Sciences, East China Normal University, Shanghai 200062, China}
\affiliation{NYU-ECNU Institute of Physics at NYU Shanghai, 3663 Zhongshan Road North, Shanghai 200062, China}
\affiliation{New York University Shanghai, 1555 Century Ave, Pudong, Shanghai 200122, China}

\author{Tim Byrnes}
\email{tim.byrnes@nyu.edu}
\affiliation{New York University Shanghai, 1555 Century Ave, Pudong, Shanghai 200122, China}  
\affiliation{State Key Laboratory of Precision Spectroscopy, School of Physical and Material Sciences, East China Normal University, Shanghai 200062, China}
\affiliation{NYU-ECNU Institute of Physics at NYU Shanghai, 3663 Zhongshan Road North, Shanghai 200062, China}
\affiliation{National Institute of Informatics, 2-1-2 Hitotsubashi, Chiyoda-ku, Tokyo 101-8430, Japan}
\affiliation{Department of Physics, New York University, New York, NY 10003, USA}

\date{\today}

\begin{abstract}
The states generated by the two-spin generalization of the two-axis countertwisting Hamiltonian are examined. We analyze the behavior at both short and long timescales, by calculating various quantities such as squeezing, spin expectation values, probability distributions, entanglement, Wigner functions, and Bell correlations. In the limit of large spin ensembles and short interaction times, the state can be described by a two-mode squeezed vacuum state; for qubits, Bell state entanglement is produced. We find that the Hamiltonian approximately produces two types of spin-EPR states, and the time evolution produces aperiodic oscillations between them. In a similar way to the basis invariance of Bell states and two-mode squeezed vacuum states, the Fock state correlations of spin-EPR states are basis invariant. We find that it is possible to violate a Bell inequality with such states, although the violation diminishes with increasing ensemble size. Effective methods to detect entanglement are also proposed, and formulas for the optimal times to enhance various properties are calculated.
\end{abstract}

\pacs{03.75.Dg, 37.25.+k, 03.75.Mn}% PACS, the Physics and Astronomy
            
\maketitle

\section{Introduction}
\label{sec:intro}

The concept of quantum squeezing has been central to the development of quantum metrology and its applications in quantum information science \cite{walls1983squeezed,scully1999quantum,gerry2005introductory,Nielsen:2011:QCQ:1972505}. 
In a squeezed state, it is possible to reduce the quantum mechanical noise (as measured by the variance) of an operator at the expense of another operator, according to the 
Heisenberg uncertainty relation. The most common experimental realizations of squeezing remain in optical systems \cite{slusher1985observation,wu1986generation,breitenbach1997measurement}. In addition to squeezed states for one mode, the two-mode squeezed state \cite{heidmann1987observation} is one of the central ingredients of optical-based quantum technologies, due to the entanglement that is possessed by this state. In a two-mode squeezed state, a linear combination of variables involving the two modes experiences suppression of quantum noise \cite{scully1999quantum,gerry2005introductory,braunstein2005quantum,ou1992realization}.  Observables in the two modes obey Einstein-Podolsky-Rosen (EPR)-like correlations, which can be employed for various quantum information tasks \cite{braunstein2005quantum}. Squeezed states of light have found applications in gravitational wave detection \cite{aasi2013enhanced}, quantum computation \cite{braunstein2005quantum}, interferometry \cite{bondurant1984squeezed}, quantum metrology \cite{giovannetti2011advances}, and quantum cryptography \cite{hillery2000quantum} to name a few examples.  

Atomic gases are another system that exhibit quantum mechanical squeezing. This includes atomic ensembles and Bose-Einstein condensates (BECs), where the relevant degrees of freedom are the internal spin states of the atoms \cite{pezze2018quantum}. In contrast to the quantum optical case where the system is described by bosonic modes, for atomic gases the appropriate description is in terms of the collective spin of all the atoms. The seminal theoretical works of Kitagawa and Ueda \cite{kitagawa1993squeezed} studied two main types of squeezed states in such atomic ensembles, the one- and two-axis spin squeezed states. These produce squeezing in the collective spin variables in a similar way to optical states. Such squeezed spin states have been observed in \cite{esteve2008squeezing,bohi2009coherent,krauter2011entanglement,zhang2014quantum}.

While squeezing on one spin has been studied in great detail theoretically and experimentally, the analogue of two-mode squeezing for the spin case is relatively less developed. The most widely known results for two spins are for atomic ensembles pioneered by the group of Polzik and co-workers \cite{Julsgaard2001,cerfbook,Krauter2013}. In these works, while the physical system is an atomic ensemble, the regime that is examined is where spin variables can be approximated by bosonic modes, according to the Holstein-Primakoff transformation. In these works the entangled state that is produced can be described within this approximation as a two-mode squeezed state.  
However, due to the fundamentally different nature of the spins to modes, it is well-known that different dynamics can result for long evolution times, observed in effects such as over-squeezing \cite{Gerving2012,Strobel2014,kajtoch2015quantum,li2017spin,wang2017two,zhang2003entanglement}. Entanglement between different spatial regions of a single BEC were experimentally observed \cite{kunkel2017,lange2017,fadel2017}. The two-spin version of one-axis squeezing was studied in Refs. \cite{byrnes2013fractality,kurkjian2013spin} where it was found that a complex structure of entanglement is present, with a time dependence showing fractal characteristics. Methods to generate this state have been examined in numerous works \cite{byrnes2013fractality,kurkjian2013spin,PhysRevA.74.022312,pyrkov2013,rosseau2014,hussain2014,jing2019split}. Procedures to generate other types of entangled states have also been studied \cite{olov,idlas2016}. Such entangled states have been studied for various applications such as quantum computing \cite{Byrnes2012,Byrnes2015} and quantum information  
\cite{hald1999spin,kuzmich2000generation,esteve2008squeezing,riedel2010atom,zhang2014quantum,krauter2011entanglement}.

In this paper, we study the two-spin version of the two-axis spin squeezed state, which we call the {\it two-axis two-spin squeezed} (2A2S) state. Within the Holstein-Primakoff approximation, corresponding to small interaction times, this state is equivalent to the two-mode squeezed state. In this approximation, the state has been studied before \cite{kupriyanov2001}. Beyond these times, the dynamics start to differ significantly from the two-mode squeezing interaction. We will particularly focus on understanding the states for longer interaction times, where the Holstein-Primakoff approximation is no longer valid. For the one-spin case, longer interaction times were studied in Refs. \cite{kajtoch2015quantum, yukawa2014}. We will generally be interested in the regime where the spin is large but finite, which is applicable for atomic gases. Even for BECs where the atomic numbers are much less than in thermal atomic ensembles, there can be easily more than $ 10^3$ atoms in an ensemble making the spins very large \cite{bohi2009coherent,riedel2010atom,pezze2018quantum}. 
We study the two-spin two-axis spin squeezed state through various quantities, calculating correlations and probability distributions (Sec. \ref{sec:corr}), entanglement (Sec. \ref{sec:quant}), Wigner functions (Sec. \ref{sec:wigner}), and Bell correlations (Sec. \ref{sec:bell}). In addition to elucidating the nature of the dynamics and the state, we interestingly find that the state can violate a Bell inequality without parity measurements.

\section{The two-axis two-spin squeezed state}
\label{sec:twoaxis}

The system that we shall consider consists of two neutral atomic ensembles or BECs, where each atom has two relevant internal states. A common choice for the internal states are hyperfine ground states, such as the $ F= 1, m_F = - 1 $ and $ F = 2, m_F = 1 $ states in the case of $^{87}$Rb
\cite{pezze2018quantum}. In the case of BECs we denote the bosonic annihilation operator for the two states as $ a_j, b_j $ respectively, where $ j \in \{1,2 \} $ labels the two BECs.  These operators can be used to define an effective spin using the Schwinger boson operators 
\begin{align}\label{eq:spinops}
    S^x_j & =b_j^\dagger a_j+a_j^\dagger b_j \nonumber \\
    S^y_j & =-ib_j^\dagger a_j+ia_j^\dagger b_j \nonumber \\
     S^z_j & =b_j^\dagger b_j-a_j^\dagger a_j  .  
\end{align}
The commutation relation for the spin operators are
\begin{align}
[S^j,S^k]=2i\epsilon_{jkl}S^l ,  \label{commutator}
\end{align}
where $\epsilon_{jkl}$ is the completely antisymmetric tensor.

For atomic ensembles, the total spin operators are written 
\begin{align}\label{eq:spinopsensemble}
    S^x_j & =  \sum_{l=1}^N \sigma^x_{j,l}  \nonumber \\
    S^y_j & =\sum_{l=1}^N \sigma^y_{j,l}  \nonumber \\
     S^z_j & =\sum_{l=1}^N \sigma^z_{j,l}  ,
\end{align}
where $ \sigma^k_{j,l} $ is a Pauli operator for the $l$th atom in the $j$th ensemble. 
For simplicity, we take the total number of atoms $N $ to be equal in both ensembles. The total spin operators also have the same commutation relation (\ref{commutator}). For an atomic ensemble where the initial state and the Hamiltonian are completely symmetric with respect to particle interchange on a single ensemble, exactly the same results are obtained with either (\ref{eq:spinops}) or (\ref{eq:spinopsensemble}). Mathematically, the BEC form (\ref{eq:spinops}) is more convenient for calculations, and hence we will use this throughout this paper. However, it should be understood that our results equally apply to the atomic ensemble case.

The two-axis two-spin (2A2S) Hamiltonian is then defined as
\begin{align}
H = H_{2A2S} = \frac{J}{2} (S_1^x S_2^x - S_1^y S_2^y) = J(S_1^+ S_2^+ + S_1^-S_2^-) ,  
\label{eq:Hamiltonian}
\end{align}
where 
\begin{align}
S_j^{+} & = \frac{1}{2} (S_j^x + i S_j^y) = b_j^\dagger a_j \nonumber \\
S_j^{-} & =  \frac{1}{2}  (S_j^x - i S_j^y)= a_j^\dagger b_j ,
\end{align}
and $ J $ is an energy constant. This is a straightforward generalization of the two-axis one-spin (2A1S) countertwisting Hamiltonian studied by Kitagawa and Ueda \cite{kitagawa1993squeezed}
\begin{align}
H_{2A1S} = \frac{J}{2} [ (S^x)^2  - (S^y)^2 ] = J[ (S^+)^2 + (S^-)^2 ] .  
\end{align}
The 2A1S Hamiltonian produces squeezing and antisqueezing in the transformed variables 
\begin{align}
\tilde{S}^x & = \frac{S^x + S^y}{\sqrt{2}} \nonumber \\
\tilde{S}^y & = \frac{S^y - S^x}{\sqrt{2}} 
\label{tildevariables}
\end{align}
respectively \cite{kitagawa1993squeezed}. We note that a similar generalization was studied previously to generalize the one-axis one-spin (1A1S) twisting Hamiltonian \cite{byrnes2020quantum}
\begin{align}
H_{1A1S} = J (S^z)^2
\end{align}
to the one-axis two-spin (1A2S) Hamiltonian \cite{byrnes2013fractality,kurkjian2013spin}
\begin{align}
H_{1A2S} = J S^z_1 S^z_2 .  
\end{align}
The benefits of the 2A1S Hamiltonian are that it can attain a higher level of squeezing than the 1A1S Hamiltonian, and that the axes for optimal squeezing are fixed \cite{kitagawa1993squeezed}. 

The 2A2S squeezed states are produced by a unitary evolution according to the Hamiltonian (\ref{eq:Hamiltonian}) for a time $ t $, 
\begin{align}
|\psi(t) \rangle & = e^{-i H t/\hbar} | 0 ,0 \rangle \rangle_1 | 0,0 \rangle \rangle_2
\nonumber \\
& = e^{-i (S_{1}^{+}S_{2}^{+}+S_{1}^{-}S_{2}^{-} ) \tau } | 0,0 \rangle \rangle_1 | 0,0 \rangle    \rangle_2 ,
\label{state}
\end{align}
where we have defined a dimensionless time $ \tau = J t/\hbar $.  The initial states are maximally polarized states in the $ S^z $-direction, the same as for the 2A1S Hamiltonian. Here we defined the spin coherent states as
\begin{align}
    | \theta , \phi \rangle\rangle_j =\frac{( \cos \frac{\theta}{2} e^{-i \phi /2 }  b_j^\dagger+ \sin \frac{\theta}{2} e^{i \phi /2 } a_j^\dagger)^N}{\sqrt{N!}}| \text{vac} \rangle ,
\end{align}
where $\theta , \phi$ are the angles on the Bloch sphere, and $ | \text{vac} \rangle $ is the vacuum containing no atoms, and $j=1,2 $ label the BECs. We also define the Fock states as 
\begin{align}
|k \rangle_j = \frac{(b_j^\dagger)^k (a_j^\dagger)^{N-k}}{\sqrt{k! (N-k)!}} | \text{vac} \rangle .
\label{fockstates}
\end{align}
The Fock states are eigenstates of the $ S^z $ operator according to
\begin{align}
S^z_j |k \rangle_j = (2k-N) |k \rangle_j  .
\end{align}

For small times we may expand the exponential in (\ref{state}) to second order, where the first terms in the expansion are
\begin{align}
|\psi(t)\rangle \approx & ( 1 - \tau^2 N^2) | N \rangle_1 | N \rangle_2 -i\tau N |N-1 \rangle_1 |N-1 \rangle_2 \nonumber \\
& - \tau^2 N (N-1)  |N-2 \rangle_1 |N-2 \rangle_2 + \dots ,
\label{expexpan}
\end{align}
where the states we have used are Fock states (\ref{fockstates}). All terms in the superposition have the same Fock numbers for the two BECs, hence we immediately observe that the state is perfectly correlated in $ S^z_j $.  We can also see that for  $ |\tau|\ll 1/N $, the population of the $ a $-state will be small.  A more precise criterion will be obtained in the next section.

In contrast to the 1A2S Hamiltonian, for which it is straightforward to find an analytical expression for the state at arbitrary evolution times, the 2A2S Hamiltonian cannot be diagonalized using a linear transformation of the bosonic operators. Only recently, static solutions for the 2A1S Hamiltonian were found \cite{Pan2017analyticsolution}. We thus resort to numerical methods to study the state and its properties. For pure state evolution, this involves evolving a vector of dimension $ (N+1)^2 $, hence reasonably large systems can still be accessed with numerical methods.

In this paper we will focus on the theoretical state that is produced by the 2A2S Hamiltonian. Although we will not discuss the experimental procedures for generating the 2A2S Hamiltonian here, we note that several possibilities exist for producing the state, where techniques to produce the 2A1S Hamiltonian can be applied to the two-spin case. Several works have considered techniques to generate the 2A1S Hamiltonian \cite{liu2011spin,huang2015two,borregaard2017one}. This can then be converted to the 2A2S Hamiltonian using a split-squeezing approach such as that given in Refs. \cite{oudot2019bell,jing2019split}. An alternative procedure is to generate the same correlations using optical interference methods \cite{Julsgaard2001,olov}. We plan to examine the experimental methods to produce the 2A2S Hamiltonian in a later work.

\section{Correlations and probability distributions} \label{sec:corr}

\subsection{Holstein-Primakoff limit}

To obtain some intuition about the state (\ref{state}), let us first examine the state for small evolution times. For a system with fixed particle number $ N $, the Holstein-Primakoff transformation \cite{holstein1940field} between spin operators to a single bosonic mode can be made according to 
\begin{align}
S^+ & = \sqrt{N - a^\dagger a} a  \nonumber \\
S^- & = a^\dagger \sqrt{N - a^\dagger a}   \nonumber \\
S^z & = N - a^\dagger a .
\end{align}
In our model, the initial state of the system is the spin coherent state $ |0, 0 \rangle \rangle $, in which all atoms in the two BECs occupy the respective internal state labeled by $b$. For $N \gg 1 $ and sufficiently small evolution times, the population of the $a$-state satisfies $ \langle a^\dagger a \rangle \ll N $. This allows us to approximate the Holstein-Primakoff-transformed spin operators as
\begin{align}
S^+ & \approx \sqrt{N} a  \nonumber \\
S^- & \approx  \sqrt{N} a^\dagger \nonumber \\
S^z & \approx  N  .  
\label{HP}
\end{align}
Note that this approximation breaks down for larger evolution times due to the finite number of atoms $N$. As the Hamiltonian acts on the system, the occupation numbers of the $a$-states increases while those of the $b$-states deplete. At a certain time, which can be explicitly seen in the subsequent sections, the condition  $ \langle a^\dagger a \rangle \ll N $ is no longer satisfied and the approximation fails.

Applying (\ref{HP}) to the 2A2S Hamiltonian, we have
\begin{align}
H_{2} \approx J N (a_1 a_2 + a_1^\dagger a_2^\dagger) . 
\label{twomodesqueezeham}
\end{align}
This is exactly the two-mode squeezing Hamiltonian \cite{scully1999quantum,gerry2005introductory,braunstein2005quantum} considered in quantum optics. The transformation of the mode operators is 
\begin{align}
e^{iH_{2} t/\hbar  } a_1 e^{-iH_{2} t/\hbar} = a_1 \cosh N \tau  - i a_2^\dagger \sinh N \tau  \nonumber \\
e^{iH_{2}t /\hbar} a_2 e^{-iH_{2} t/\hbar} = a_2 \cosh N \tau  - i a_1^\dagger \sinh N \tau .
\end{align}

We can deduce the time for which the Holstein-Primakoff approximation is valid by evaluating the population of the $a_1$ and $a_2$ states.  The population of the two states are always equal $ \langle a_1^{\dagger}  a_1 \rangle = \langle a_2^{\dagger}  a_2 \rangle  $ and we obtain
\begin{align}
\langle a_1^{\dagger}(t)  a_1(t) \rangle &= \langle 0 | (a_1^\dagger \cosh N \tau  + i a_2 \sinh N \tau) \nonumber \\ 
&\times (a_1 \cosh N \tau  - i a_2^\dagger \sinh N \tau) | 0\rangle  \nonumber \\ 
&= \sinh ^2 N\tau \approx e^{2N\tau}/4 ,
\end{align}
where in the last step we assumed $ N\tau \gg 1 $.  Demanding that $ \langle a_1^\dagger a_1 \rangle \ll N $, we have the criterion for the validity of the Holstein-Primakoff approximation as
\begin{align}
\tau \ll \frac{\ln(4N)}{2N} .
\label{hpvalidtime}
\end{align}

Now let us define the canonical position and momentum operators as 
\begin{align}
x_j & = \frac{a_j + a^\dagger_j}{\sqrt{2}} \approx \frac{S^+_j + S^-_j}{\sqrt{2N}} = \frac{S^x_j}{\sqrt{2N}}  \nonumber \\
p_j & = \frac{-i a_j + i a^\dagger_j}{\sqrt{2}}  \approx \frac{-i S^+_j + i S^-_j}{\sqrt{2N}}  = \frac{S^y_j}{\sqrt{2N}} . 
\end{align}
For the choice of phase between the two terms in (\ref{twomodesqueezeham}), the relevant operators are those that are rotated by $45^\circ$ with respect to the quadrature axes
\begin{align}
\tilde{x}_j & = \frac{x_j + p_j}{\sqrt{2}} \approx \frac{\tilde{S}^x_j}{\sqrt{2N}} \nonumber \\
\tilde{p}_j & = \frac{p_j-x_j}{\sqrt{2}} \approx \frac{\tilde{S}^y_j}{\sqrt{2N}} ,
\end{align}
where we used the definitions (\ref{tildevariables}). The correlations for which the quantum noise is suppressed are then $ \tilde{x}_1 + \tilde{x}_2  $ and $ \tilde{p}_1 - \tilde{p}_2 $  \cite{vaidman1994teleportation,braunstein2005quantum}. This can be seen by evaluating
\begin{align}
e^{i H_{2} t/\hbar} (\tilde{x}_1 + \tilde{x}_2) e^{- iH_{2} t/\hbar} & = e^{-N \tau} (\tilde{x}_1 + \tilde{x}_2) \nonumber \\
e^{i H_{2} t/\hbar} (\tilde{p}_1 - \tilde{p}_2) e^{- iH_{2} t/\hbar} & = e^{-N \tau} (\tilde{p}_1 - \tilde{p}_2) , 
\label{cvlimitsqueezed}
\end{align}
which become suppressed for large squeezing times. The corresponding anti-squeezed variables are
\begin{align}
e^{i H_{2} t/\hbar} (\tilde{x}_1 - \tilde{x}_2) e^{- iH_{2} t/\hbar} & = e^{N \tau} (\tilde{x}_1 - \tilde{x}_2) \nonumber \\
e^{i H_{2} t/\hbar} (\tilde{p}_1 + \tilde{p}_2) e^{- iH_{2} t/\hbar} & = e^{N \tau} (\tilde{p}_1 + \tilde{p}_2) .
\label{cvlimitantisqueezed}
\end{align}

\subsection{EPR-like correlations}

We now directly evaluate the correlations produced by the 2A2S Hamiltonian numerically, without applying the Holstein-Primakoff approximation. From (\ref{cvlimitsqueezed}) we expect that the variances of the observables 
\begin{align}
O_{\text{sq}} \in \{ \tilde{S}^x_1 + \tilde{S}^x_2, \tilde{S}^y_1 - \tilde{S}^y_2 \} 
\label{eprquants}
\end{align}
become suppressed, for short times when the Holstein-Primakoff approximation holds. The observables in the perpendicular directions (\ref{cvlimitantisqueezed}) 
\begin{align}
O_{\text{asq}} \in \{  \tilde{S}^x_1 - \tilde{S}^x_2, \tilde{S}^y_1 + \tilde{S}^y_2 \}
\label{eprquantsanti}
\end{align}
are the anti-squeezed variables.   

\begin{figure}[t]%
\includegraphics[width=\linewidth]{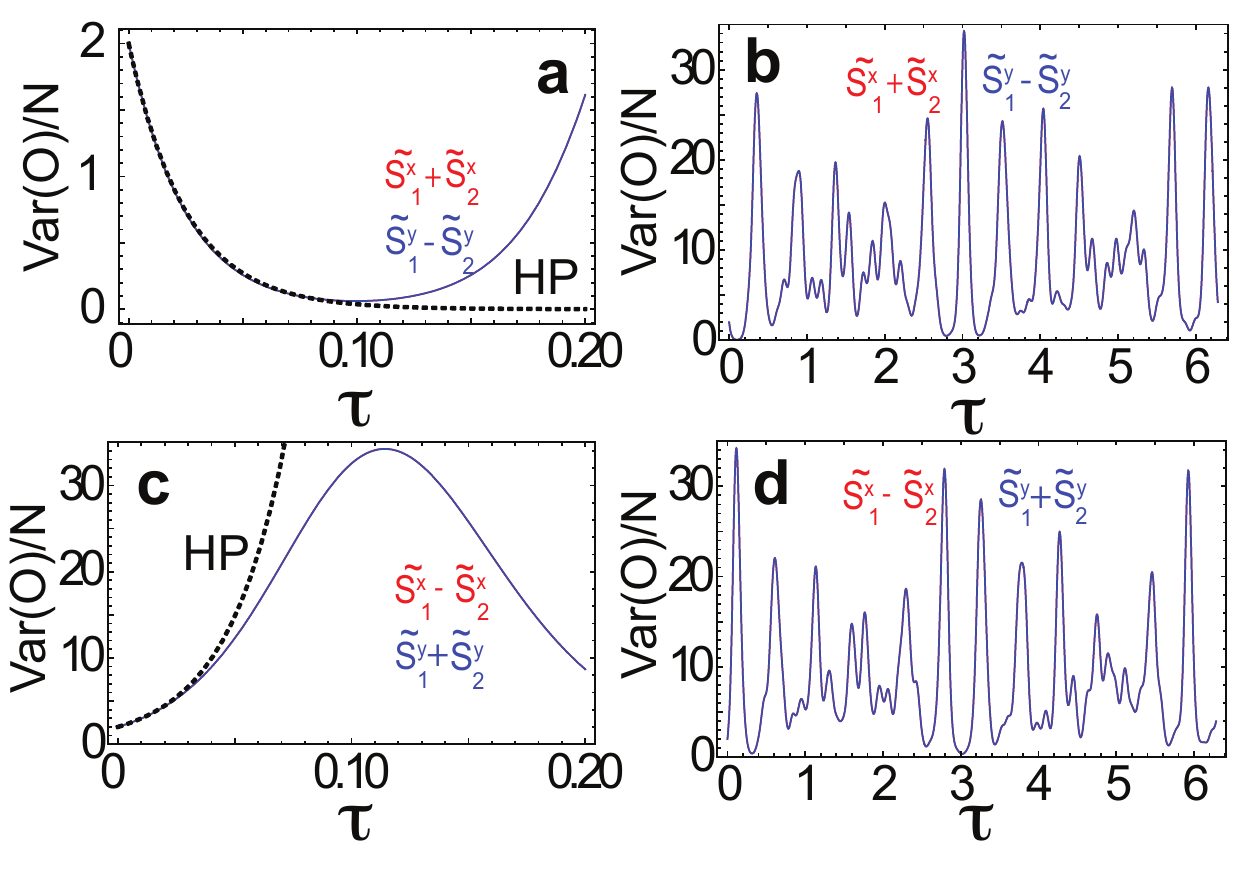}
\caption{Variances of EPR-like observables in the two-axis two-spin squeezed state.  The variances of the (a)(b) squeezed variables $ \tilde{S}^{x}_{1}+ \tilde{S}^{x}_{2}$ and $ \tilde{S}^{y}_{1}- \tilde{S}^{y}_{2}$; (c)(d) anti-squeezed variables $ \tilde{S}^{x}_{1}- \tilde{S}^{x}_{2}$ and $ \tilde{S}^{y}_{1}+ \tilde{S}^{y}_{2}$ are plotted as a function of the dimensionless interaction time $ \tau $.  The Holstein-Primakoff (HP) approximated variances are shown by the dotted lines. (a)(c) show timescales in the range $ \tau \sim 1/N $ and (b)(d) show longer timescales $ \tau \sim 1 $.  The number of atoms per ensembles is taken as  $N = 20$. }
\label{fig1}%
\end{figure}

In Fig. \ref{fig1}(a), the variances of the observables (\ref{eprquants}) are plotted for short timescales  $\tau \sim 1/N $. We see that the two variances have exactly the same time dependence and take a minimum at a time which we define as the optimal squeezing time $ \tau_{\text{opt}}^{(\text{sq})} $. 
In the time region $ 0 \le \tau \le \tau_{\text{opt}}^{(\text{sq})}  $, the variance agrees well with Holstein-Primakoff approximation, giving
\begin{align}
\text{Var} (O_{\text{sq}}, \tau ) & \approx 2N e^{-2N \tau},
\label{squeezedvarhp}
\end{align}
which follows from the relations (\ref{cvlimitsqueezed}) and the fact that $ \text{Var} (O_{\text{sq}}, \tau=0 ) = 2N $. Beyond these times the variance increases and no longer follows (\ref{squeezedvarhp}).  For longer timescales, as shown in Fig. \ref{fig1}(b), the variance follows aperiodic oscillations between low and high variance states.  Some relatively low variance states are achieved (e.g., particularly around $ \tau \approx 3 $), although the minimum variance at the times $ \tau_{\text{opt}}^{(\text{sq})} $ is not attained again.

The anti-squeezed variables (\ref{eprquantsanti}) are shown in  Fig. \ref{fig1}(c) for short timescales  $  \tau \sim 1/N $. Again the two variables (\ref{eprquantsanti}) have exactly the same time dependence, and initially increase according to
\begin{align}
\text{Var} (O_{\text{asq}}, \tau ) & \approx 2N e^{2N\tau},
\end{align}
which follows from (\ref{cvlimitantisqueezed}).  In contrast to genuine two-mode squeezing, the variance does not increase unboundedly but reaches a maximum. We call this time the optimal anti-squeezing time  $ \tau_{\text{opt}}^{(\text{asq})} $, which we find is not exactly at the same time as the optimal squeezing time 
$ \tau_{\text{opt}}^{(\text{sq})} $.  As with the squeezed variables, for longer timescales, as shown in Fig. \ref{fig1}(d), the anti-squeezed variables show aperiodic oscillations, with a similar range to Fig. \ref{fig1}(b). Some low variances states are also present with the anti-squeezed variables, again around $ \tau \approx 3 $.  However, these states do not reach the low level of the variance attained with the squeezed variables (\ref{eprquants}).  

The existence of a bound on the amount of squeezing, as opposed to unbounded genuine two-mode squeezing, is a consequence of the finite atom number $N$. In the large $N$ limit, the Holstein-Primakoff approximation is valid and our system is equivalent to two-mode squeezing. Restricting the atom number $N$ to a finite value renders the Fock space finite-dimensional and leads to oscillations of the populations between the north and south poles of the Bloch sphere, as will be seen in the following section.

 \begin{figure}[ht]%
\includegraphics[width=\linewidth]{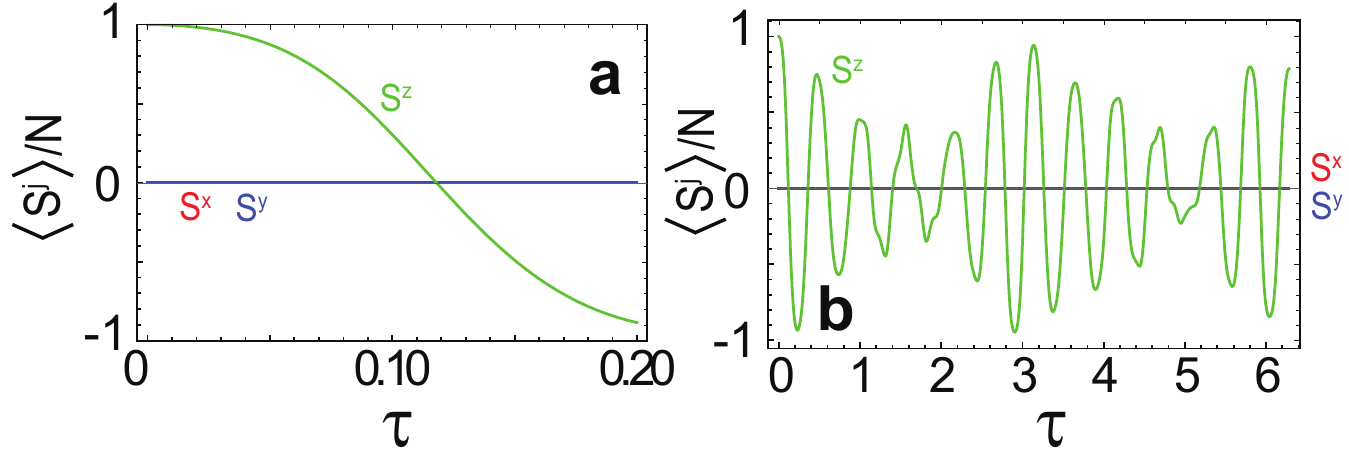}
\caption{Expectation values of spin operators for the two-axis two-spin squeezed state for (a) short timescales $ \tau \sim 1/N $; (b) long timescales  $ \tau \sim 1$.  The number of atoms per ensembles is taken as  $N = 20$. }
\label{fig2}%
\end{figure}

\subsection{Expectation values}

It is also instructive to examine the expectation values of the spin operators in the 2A2S squeezed state.  Fig. \ref{fig2} shows the expectation values of the operators $ S^x_j$, $S^y_j$, $ S^z_j $.  Due to the symmetry between the initial state of the two ensembles and the 2A2S Hamiltonian, identical values are obtained for the two ensembles $ j = 1,2 $.  Furthermore, the expectation values of two of the operators are always zero:
\begin{align}
\langle S^x_j (\tau) \rangle = \langle S^y_j (\tau) \rangle = 0  .
\end{align}
This can be seen from (\ref{expexpan}), where the Hamiltonian creates pairs of equal number Fock states.  Since the $S^x $ and $S^y $ operators shift the Fock states by one unit, 
\begin{align}
S^x_j |k \rangle_j &=\sqrt{(N-k)(k+1)} |k+1 \rangle _j \nonumber \\
&+ \sqrt{(N-k+1)k} |k-1 \rangle _j  \nonumber \\
S^y_j |k \rangle_j &= -i\sqrt{(N-k)(k+1)} |k+1 \rangle _j \nonumber  \\
&+ i\sqrt{(N-k+1)k} |k-1 \rangle _j .
\end{align}
the expectation values of $ S^x $ and $ S^y $ are zero for all time. 

Meanwhile, the expectation value of  $ S^z_j $ undergoes aperiodic oscillations and flips sign numerous times during the evolution. In particular, a sign change is observed in the vicinity of $ \tau_{\text{opt}}^{(\text{asq})} $. This can be understood from (\ref{expexpan}), where at $ \tau \sim 1/N $ the sum contains all terms with a similar magnitude. Again, the time where $ \langle S^z_j \rangle = 0 $ is not exactly the same as the optimal squeezing time $ \tau_{\text{opt}}^{(\text{sq})} $ or anti-squeezing time $ \tau_{\text{opt}}^{(\text{asq})} $.  We label the first time that a sign flip of $ S^z_j $ occurs by $ \tau_{\text{opt}}^{(S^z)} $. We may thus picture the type of state that is produced as a two-spin version of the planar squeezed state, where the squeezing is observed in the $ (S^x_j, S^y_j) $ plane \cite{He_2012}.

\subsection{Optimal squeezing times}

\begin{figure}[t]
\includegraphics[width=\linewidth]{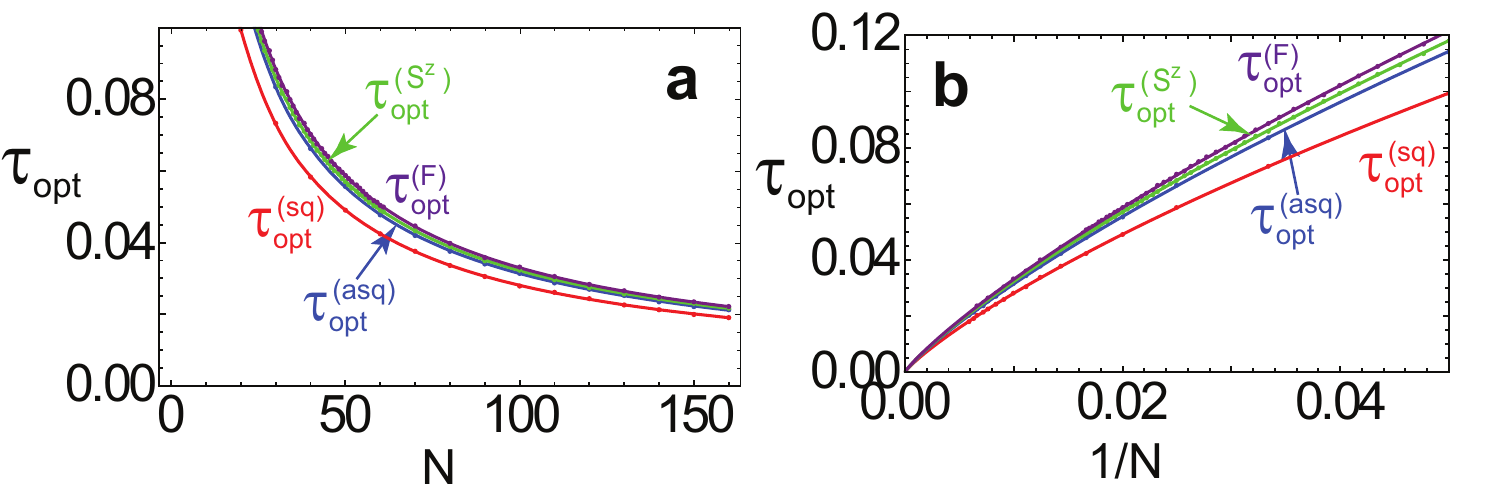}
\caption{Optimal squeezing times $ \tau_{\text{opt}}^{(\text{sq,asq},S^z,F)} $ as extracted from minimizing  $ \text{Var} (\tilde{S}^{x}_{1}+ \tilde{S}^{x}_{2} ) $,  maximizing $  \text{Var} ( \tilde{S}^{x}_{1} - \tilde{S}^{x}_{2} )$, finding the first zero of  $ \langle S^z  \rangle $, 
and optimizing the fidelity of the spin-EPR state (\ref{fidelityspinepr}), respectively.  (a) is plotted against $ N $ and (b) is plotted against $ 1/N $ for the same data.  Points show the numerically obtained values, lines are fit lines using (a) spline interpolation
(b) the fit function (\ref{newfitform}).  The fit parameters for minimized squeezing are $ p_0 = 0.467, p_1 = 0.508 $, maximum anti-squeezing are $ p_0 = 0.700, p_1 = 0.530 $, zeros of $ \langle S^z \rangle $ are $ p_0 = 0.727, p_1 = 0.536 $, the fidelity with the spin-EPR state are $ p_0 = 0.803, p_1 = 0.544 $.
 }
\label{fig3}%
\end{figure}

In order to maximize the correlations between the two spins, it would be useful to have a general expression that gives the optimal squeezing time $\tau_{\text{opt}}^{(\text{sq})} $.  Unfortunately, we have not been able to find an analytical expression to give this for general $ N $.  However, it is possible to find approximate formulas that can give this to good accuracy.  We will investigate this in the following. 

As can be seen from Fig. \ref{fig1}(a)(c) and Fig. \ref{fig2}(a), the optimal squeezing time $\tau_{\text{opt}}^{(\text{sq})} $, optimal anti-squeezing time  $\tau_{\text{opt}}^{(\text{asq})} $, and the time $\tau_{\text{opt}}^{(S^z)} $ of the first zero of $ \langle S^z \rangle $ do not necessarily coincide. The optimal times according to each criterion are shown in Fig. \ref{fig3}(a). Generally, the optimal squeezing time tends to give slightly smaller values than that for anti-squeezing, or the time for the first zero of $ \langle S^z \rangle$.  The latter two conditions give very similar values of the optimal time.

In Fig. \ref{fig1}(a) we notice that the time that the Holstein-Primakoff approximation starts to deviate from the exact expression coincides with the optimal time for squeezing. This suggests that the functional form of (\ref{hpvalidtime}) may be a suitable form to fit the optimal times in Fig. \ref{fig3}(a).  Using the fitting form
\begin{align}
\tau_{\text{opt}} \approx \frac{p_0 + p_1 \ln N }{N}
\label{newfitform}
\end{align}
we fit the various optimal times as shown in Fig. \ref{fig3}(b).  This fitting function works very well, with the logarithmic term accounting for the non-linear behavior seen in the $ 1/N $ plot. For the optimal time extracted for the squeezed variables we obtain fit parameters $ p_0 = 0.467, p_1 = 0.508 $, close to the theoretically calculated values of (\ref{hpvalidtime}), $ p_0 = \ln 2, p_1 = 1/2 $.   Since (\ref{newfitform}) is guaranteed to approach $ \tau_{\text{opt}}  \rightarrow 0 $ in the limit of large $ N $, we expect that this fit will interpolate the larger $N $ optimal times with good accuracy. The parameters for the other optimal times are given in the caption of Fig. \ref{fig3}.

Using the optimal squeezing times, the maximal squeezing that can be attained can be estimated from the Holstein-Primakoff relation according to 
\begin{align}
\min_\tau  \text{Var} (\tilde{S}^x_1 + \tilde{S}^x_2, \tau) & = 
\min_\tau  \text{Var} (\tilde{S}^y_1 - \tilde{S}^y_2, \tau)  \nonumber \\
& \approx 2Ne^{-2N \tau_{\text{opt}}^{(\text{sq})} } \nonumber \\
& \approx  \frac{2 N}{e^{2p_0} N^{2 p_1}} , 
\label{minvaluesqueezed}
\end{align}
where in the last line we substituted the expression (\ref{newfitform}). The minimal squeezing level tends to improve approximately as $ 1/N $ for larger ensemble sizes due to the factor of $ N^{2 p_1} $ in the denominator, relative to the spin coherent state variance $ 2N $.

\subsection{Probability density of the two-axis co-squeezed state}

Another way to visualize the correlations is to plot the probability distributions when the state (\ref{state}) is measured in various bases. Specifically we consider the Fock states which are the eigenstates of the rotated operators (\ref{tildevariables})
\begin{align}
\tilde{S}^x | k \rangle^{(\tilde{x})} &  = (2 k - N ) | k \rangle^{(\tilde{x})}  \nonumber \\
\tilde{S}^y | k \rangle^{(\tilde{y})} &  = (2 k - N ) | k \rangle^{(\tilde{y})}  \nonumber \\
S^z | k  \rangle^{(z)} &  = (2 k  - N ) | k \rangle^{(z)} .
\end{align}
These Fock states can be transformed from the $S^z $-eigenstate Fock states (\ref{fockstates}) using the relations in Appendix \ref{app:prob}. The probability of a measurement outcome $ k_1, k_2 $ for various Fock states is then 
\begin{align}
p_{l_1 l_2} (k_1, k_2 ) = | \langle \psi (t) | 
\left[ |k \rangle^{(l_1)} \otimes | k \rangle^{(l_2)}  \right]  |^2 ,
\end{align}
where $ l_1, l_2 \in \{ x,y,z \} $.  

\begin{figure}
\includegraphics[width=\columnwidth]{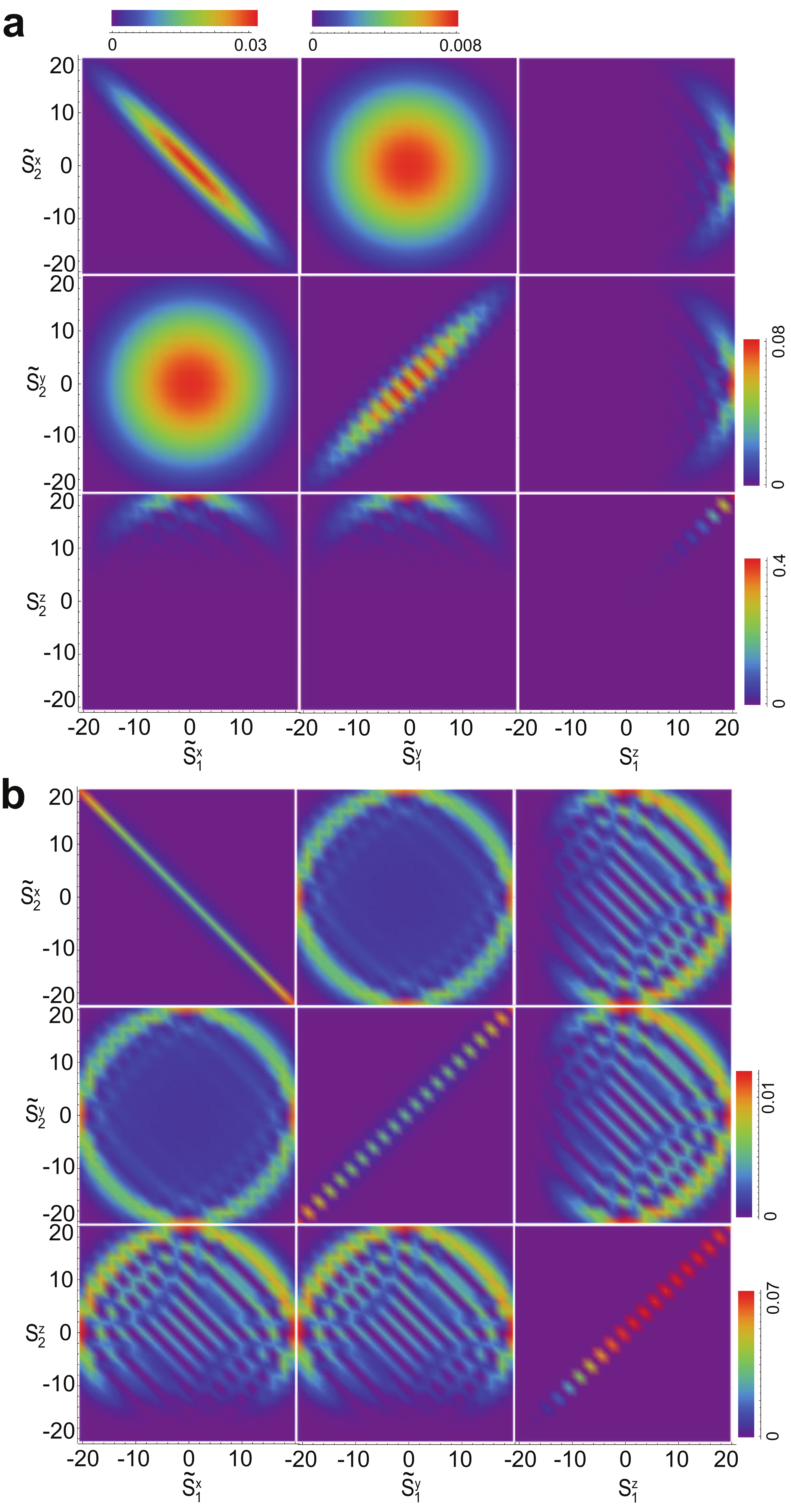}
\caption{Probability distributions of the two-axis two-spin state measured in various bases for (a) $ \tau = 0.05 \approx \tau_{\text{opt}}^{(\text{sq})} /2 $; (b) $ \tau = 0.1 \approx \tau_{\text{opt}}^{(\text{sq})}  $. The density plot legends for similar shaped distributions are the same.  In (a), the legend for the $ (\tilde{S}^y_1,\tilde{S}^y_2) $ basis is the same as $ (\tilde{S}^x_1,\tilde{S}^x_2 ) $. In (b), the legend for the 
$ (\tilde{S}^y_1,\tilde{S}^x_2) $ and $ (S^z_1,\tilde{S}^x_2) $ is the same as $ (S^z_1,\tilde{S}^y_2 )$. The total number of atoms per BEC is $ N = 20 $ for all plots.}
\label{fig4}
\end{figure}

The probabilities for two evolution times before and near $\tau_{\text{opt}}^{(\text{sq})} $ are shown in Fig. \ref{fig4}. The effect of the correlations are seen in the $ (\tilde{S}^x_1, \tilde{S}^x_2 )$ and $ (\tilde{S}^y_1, \tilde{S}^y_2 )$ measurement combinations, where the most likely probabilities occur when $ \tilde{S}^x_1 = -\tilde{S}^x_2 $ and $  \tilde{S}^y_1 = \tilde{S}^y_2 $ respectively. This means that the quantities $ \tilde{S}^x_1 + \tilde{S}^x_2 $ and $  \tilde{S}^y_1 - \tilde{S}^y_2 $ always take small values and hence are squeezed. The probability distribution for the measurements for the four cases $ (\tilde{S}^{x,y}_1, \tilde{S}^{x,y}_2 )$ initially starts as a Gaussian centered around $ \tilde{S}^{x,y}   = 0 $ and becomes increasingly squeezed.  For the  $ (S^z_1, S^z_2 )$ measurement, we see Fock state correlations arising from the fact that the 2A2S Hamiltonian always produces Fock states in pairs, as shown in (\ref{expexpan}).  

Comparing the $ (\tilde{S}^x_1, \tilde{S}^x_2 )$ and $ (\tilde{S}^y_1, \tilde{S}^y_2 )$ in more detail reveals some interesting effects. For the sub-optimal squeezing time in  Fig. \ref{fig4}(a) we see that the peak of the distribution is at $ \tilde{S}^x_1 = \tilde{S}^x_2 = 0 $, which is the original position of the Gaussian before the Hamiltonian is applied.  In Fig. \ref{fig4}(b), we see that the maximum is at the ends of the distribution. Thus the optimal squeezing corresponds approximately to tuning the time such that the $ (\tilde{S}^x_1, \tilde{S}^x_2 )$ and $ (\tilde{S}^y_1, \tilde{S}^y_2 )$ distributions are at their flattest.

For the remaining correlation pairs, the distributions are always symmetrical in the variables $ \tilde{S}^x $ and $ \tilde{S}^y $, hence give zero when averaged. Thus there is no correlation between the remaining variables. The lack of correlations in the off-diagonal combinations in Fig. \ref{fig4} is also a feature of standard Bell states. The type of state that is considered here is therefore a natural generalization of the EPR correlated state for spin ensembles, which we discuss further in Sec. \ref{sec:spinepr}.

\section{Entanglement} \label{sec:quant}

\subsection{Time dependence of entanglement for pure states}

We now turn to the entanglement that is generated in 2A2S state. The entanglement that we consider is that present between the two BECs, which forms a natural bipartition in the system. We will not consider other types of intra-ensemble entanglement here, as have been considered for single ensemble squeezed states \cite{pezze2018quantum}.  For the 2A2S squeezed state, there is no squeezing on a single BEC, hence we do not expect intra-ensemble entanglement to be a relevant quantity in this context.

\begin{figure}
\includegraphics[width=\columnwidth]{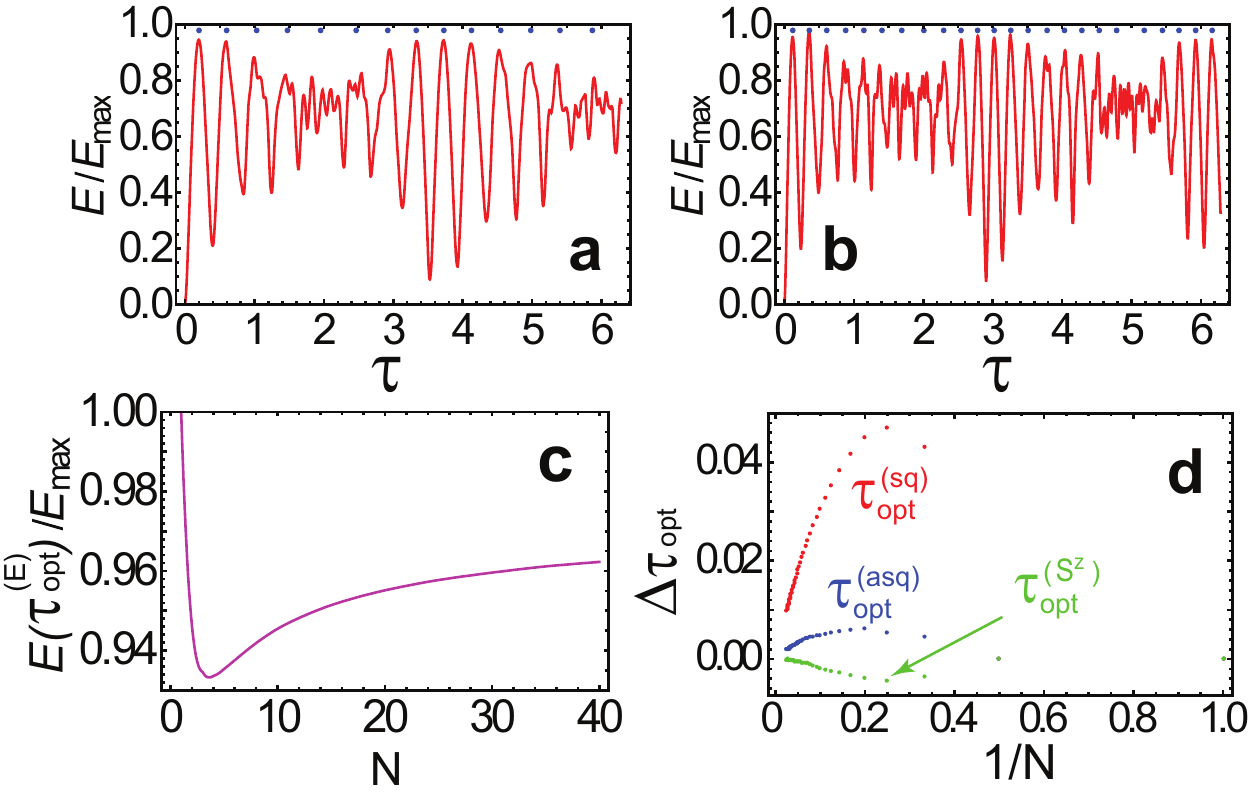}
\caption{Entanglement of the two-axis two-spin squeezed state, as measured by the von Neumann entropy $E$  normalized to the maximum value $E_{\text{max}} = \text{log}_2 (N+1)$, as a function of the interaction time $\tau$. The number of atoms in each BEC is (a) $N = 10$; (b) $ N = 20 $. The dots in each figure denote the times when $ \langle S^z_1 \rangle = 0 $ for each $N $.  (c) The maximum value of the von Neumann entropy as a function of $ N $. (d) Difference in the optimal time $ \Delta \tau_{\text{opt}} = \tau_{\text{opt}}^{(\text{E})} - \tau_{\text{opt}}^{(\text{sq,asq,$S^z$})} $, where $ \tau_{\text{opt}}^{(\text{E})} $ is the optimal time according to von Neumann entropy, $ \tau_{\text{opt}}^{(\text{sq})} $ is for the squeezed variables, $ \tau_{\text{opt}}^{(\text{asq})} $ is for the anti-squeezed variables, $ \tau_{\text{opt}}^{(S^z)} $ is for zero of $ \langle S^z \rangle $.  \label{fig5} }
\end{figure}

For pure states, bipartite entanglement can be quantified using the von Neumann entropy 
\begin{align}
E (t)= -\Tr\left( \rho_2 \text{log}_2 \rho_2 \right),
\end{align}
where 
\begin{align}
\rho_2(\tau)  = \Tr_1 ( | \psi (\tau) \rangle \langle \psi (\tau) |) 
\label{eq:rhotraced}
\end{align}
is the reduced density matrix for BEC 2. Figure \ref{fig5}(a)(b) shows the von Neumann entropy normalized to the maximum value $E_{\text{max}} = \text{log}_2(N+1)$ for two $N+1$ level systems.  We see that the entanglement first reaches a maximum at a similar time to the optimal squeezing time $ \tau_{\text{opt}}^{(\text{sq})} $, and reaches nearly the maximum possible entanglement between the two BECs.  For larger values of $N $, the oscillations have a higher frequency, with a period that is $ \sim 2 \tau_{\text{opt}} $.  

Figure \ref{fig5}(c) shows the maximal entanglement as a function of $ N $.  For each $ N $, we find the maximum value of the entanglement by optimizing the time in the vicinity of the first maximum. We call the time that this occurs the optimal entanglement time $ \tau_{\text{opt}}^{(E)} $.  We see that the optimized entanglement approaches the maximum possible entanglement $E_{\text{max}} $ for large $N$. This is in agreement with estimates from the Holstein-Primakoff approximated Hamiltonian, where the maximum value is reached for $N \rightarrow \infty $. The convergence to the maximum value however does occur logarithmically (see Appendix \ref{app:hpentanglement}), as observed by the slow approach of Fig. \ref{fig5}(c). 
The entanglement oscillates between large and small values and tends to occur at the values corresponding to $ \langle S^z \rangle = 0 $. This is true not only in the vicinity of the first maximum that is reached at $ \tau_{\text{opt}}^{(S^z)} $, but for all $ \tau $.  Figure \ref{fig5}(a)(b) marks all the times (with a dot in the figure) where $ \langle S^z \rangle =0 $. We see that each peak in the entanglement occurs when $ \langle S^z \rangle =0 $. This is reasonable to expect from the point of view that zero values of $ \langle S^z \rangle $ correspond to states that potentially have large degrees of correlation in the 
$ \langle S^x \rangle $ and $ \langle S^y \rangle $ variables. In Fig. \ref{fig5}(d) we compare the various optimal squeezing times to the time to maximize the entanglement $ \tau_{\text{opt}}^{(E)} $.  The time that most closely approximates the maxmimal entanglement is $ \tau_{\text{opt}}^{(S^z)} $, which slightly overestimates the optimal entangling time, but gives the closest approximation. We point out that $ \langle S^z \rangle = 0 $ of course does not ensure that entanglement is present, particularly for mixed states, since this can equally occur for dephased states. However, this is a convenient heuristic that could be easily measured that coincides with large values of entanglement.

\subsection{The spin-EPR state}
\label{sec:spinepr}

We have seen in Fig. \ref{fig5}(c) that near-maximal entanglement can be obtained at optimized evolution times of the 2A2S Hamiltonian.  We have also seen in Fig. \ref{fig4}(b) that at the optimized squeezing times, very flat distributions of the correlations can be obtained.  
These facts suggest that a good approximation for the state in the large $N $ regime is
\begin{align}
|\psi (\tau_{\text{opt}}) \rangle & \approx | \text{EPR}_{-} \rangle ,
\end{align}
where we defined the state
\begin{align}
| \text{EPR}_{-} \rangle = \frac{1}{\sqrt{N+1}} \sum_{k=0}^N | k \rangle_1^{(\tilde{x})} | N-k \rangle_2^{(\tilde{x})}  .
\label{spinEPRxbasis}
\end{align}
This state has the maximum possible entanglement $E_{\text{max}} $ between the two BECs, and exhibits squeezing in the variable $ \tilde{S}^x_1 + \tilde{S}^x_2 $. Algebraic manipulation allows one to rewrite this state equally as
\begin{align}
| \text{EPR}_{-} \rangle & = \frac{1}{\sqrt{N+1}} \sum_{k=0}^N (-1)^k | k \rangle_1^{(\tilde{y})} | k \rangle_2^{(\tilde{y})}  \label{spinEPRybasis}  \\
& = \frac{1}{\sqrt{N+1}} \sum_{k=0}^N (-1)^k | k \rangle_1^{(z)} | k \rangle_2^{(z)}  ,
\label{spinEPRzbasis}
\end{align}
which have the correct $ \tilde{S}^y_1 - \tilde{S}^y_2 $ and $ \tilde{S}^z_1 - \tilde{S}^z_2 $ correlations, in agreement with Fig. \ref{fig4}. Such a state is a type of spin-EPR state which exhibits correlations in a similar way to Bell states and continuous variable two-mode squeezed states, in all possible bases \cite{braunstein2005quantum}.  In fact, as we show in Appendix \ref{app:epr}, the correlations are for any choice of basis such that 
\begin{align}
| \text{EPR}_{-} \rangle  = \frac{1}{\sqrt{N+1}} \sum_{k=0}^N  | k \rangle^{(\theta,\phi)}_1 | k \rangle^{(\theta,\pi - \phi)}_2 ,
\label{prototypeepr}
\end{align}
where the Fock states rotated by a polar $ \theta $ and azimuthal $ \phi $ angles are defined by 
\begin{align}
| k \rangle^{(\theta, \phi)} = e^{-i S^z \phi/2} e^{-i \tilde{S}^y \theta/2}  | k \rangle^{(z)} .  
\label{thetaphifock}
\end{align}

\begin{figure}
\includegraphics[width=\columnwidth]{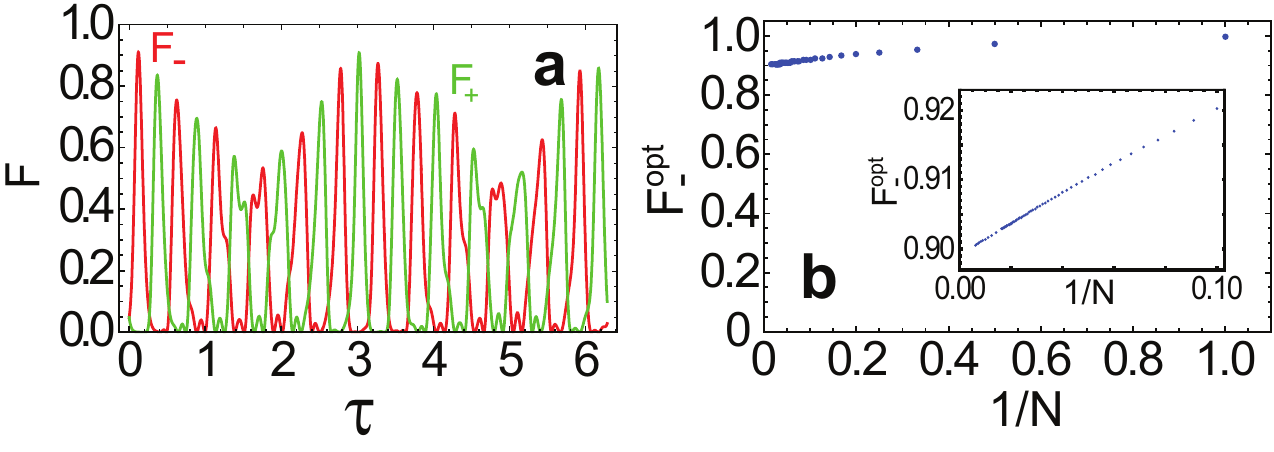}
\caption{(a) Fidelities of the two-axis two-spin squeezed state (\ref{state}) with respect to the spin-EPR states (\ref{spinEPRzbasis}) and (\ref{eprplus}).   Time dependence of the fidelity for $ N = 20 $.  (b) Optimized fidelity (\ref{spinEPRzbasis}) as a function of $ 1/N$.  Inset shows zoomed in region for large $ N $. \label{fig6}}
\end{figure}

In Fig. \ref{fig6} show the fidelity of the 2A2S squeezed state with reference to the spin-EPR state, defined as 
\begin{align}
F_{-} = | \langle \text{EPR}_{-} | \psi(\tau) \rangle |^2 .  
\label{fidelityspinepr}
\end{align}
We also plot the fidelity with respect to another spin-EPR state defined without phases
\begin{align}
F_+ = | \langle \text{EPR}_{+} | \psi(\tau) \rangle |^2 ,  
\end{align}
where 
\begin{align}
| \text{EPR}_{+} \rangle = \frac{1}{\sqrt{N+1}} \sum_{k=0}^N | k \rangle_1^{(z)} | k \rangle_2^{(z)} .  
\label{eprplus}
\end{align}
We see that the state attains high overlap with the $ | \text{EPR}_{-} \rangle  $ at a time $ \tau_{\text{opt}}^{(E)} $ as expected, and oscillates with peaks at similar times as the peaks in the anti-squeezing parameters seen in Fig. \ref{fig1}(d). On comparison with Fig. \ref{fig5}(b), we see that every second peak in the entanglement corresponds to the peaks for the fidelity $ F_- $.  The remaining peaks occur for the fidelity $F_+ $.  The timing of the peaks in $F_+ $ also match the peaks in squeezing parameters in Fig. \ref{fig1}(b). This makes it clear that the effect of the 2A2S Hamiltonian is to first generate a state closely approximating $ | \text{EPR}_{-} \rangle  $, which 
then subsequently evolves to $ | \text{EPR}_{+} \rangle  $, and this cycle repeats itself in an aperiodic fashion.  

In Fig. \ref{fig6}(b) we examine the scaling of the fidelity with $ N $.  We optimize the interaction time $ \tau $ such as to maximize $ F_- $ in the region of the optimal squeezing time. The time where the first maximum in $ F_- $ is attained is defined as $ \tau_{\text{opt}}^{(F)} $.  
The optimal time $ \tau_{\text{opt}}^{(F)} $ is again found to be most similar to $ \tau_{\text{opt}}^{(S^z)} $, but not precisely the same (see Fig. \ref{fig3}). The fidelity approaches $ F_- \approx 0.9 $ for the largest ensemble sizes that we examined ($ N = 160 $).  While the data suggests a $\propto 1/N $ relation, we do not perform an extrapolation for large $ N $, since we cannot completely rule out that logarithmic corrections may exist due to the fidelity being a sensitive quantity for large systems.   It does however appear that the dependence with $ N $ is rather weak and 
it is not unreasonable to expect that high fidelity with spin-EPR states can be generated even for realistic BEC sizes.

\subsection{Entanglement detection}

The results of Fig. \ref{fig5} clearly show that the 2A2S squeezed state is entangled for all times except for $ \tau = 0 $.  For pure states, the von Neumann entropy completely quantifies the amount of entanglement in a bipartite system. However, calculation of the von Neumann entropy relies on the availability of the complete wavefunction $ | \psi (t ) \rangle $, which may be difficult to extract experimentally, particularly for large dimensional systems. Tomographic reconstruction of the full density matrix has a high overhead in terms of the number of measurements that need to be made. Another potential experimental constraint is that only particular types of measurements may be feasible. The most common type of measurement in the context of BECs are Fock state measurements. However, at present the atom number resolution is a limitation, hence quantities that are insensitive to single atom fluctuations are preferred. Thus the experimentally preferred quantities are low order spin expectation values. In this section, we discuss criteria which only involve low order spin expectation values and can show the presence of entanglement between the two BECs.  

We compare three potential criteria that can be used to detect entanglement using low order spin expectation values.  
The first is the Giovannetti-Mancini-Vitali-Tombesi (GMVT) criterion \cite{giovannetti2002entanglement}, which states that any separable state obeys
\begin{align}
\frac{\sqrt{ \text{Var}( g_x \tilde{S}^x_1 +  \tilde{S}^x_2) \text{Var}( g_y  \tilde{S}^y_1 -  \tilde{S}^y_2) }}{ | g_x g_y| 
( |\langle S^z_1 \rangle |+  |\langle S^z_2 \rangle| ) } \ge 1  .  
\label{eq:Giovannetti}
\end{align}
Here, $ g_x , g_y $ are parameters to be optimized, and we have chosen operators combinations that attain small values near the optimal squeezing times $ \tau_{\text{opt}}^{(\text{sq})} $.  We find that the optimum values of the parameters are in this case $ g_x = g_y =1 $. The second criterion is the 
Duan-Giedke-Cirac-Zoller (DGCZ) criterion \cite{Duan}, which for the type of correlations that are present in this state reads
\begin{align}
\frac{\text{Var}( \tilde{S}^x_1 +  \tilde{S}^x_2) + \text{Var}( \tilde{S}^y_1 -  \tilde{S}^y_2)}{
2( |\langle S^z_1 \rangle |+  |\langle S^z_2 \rangle| ) } \ge 1 ,
\label{eq:duan}
\end{align}
which is true for any separable state.  The third criterion is the Hofmann-Takeuchi (HT) criterion \cite{PhysRevA.68.032103} based on local uncertainty relations for three spin operators
\begin{align}
\frac{\text{Var}( \tilde{S}^x_1 +  \tilde{S}^x_2) + \text{Var}( \tilde{S}^y_1 - \tilde{S}^y_2) 
+ \text{Var}( S^z_1 - S^z_2)}{
4N } \ge 1 ,
\label{eq:hofmann}
\end{align}
which is true for any separable state. Violation of the inequalities (\ref{eq:Giovannetti}), (\ref{eq:duan}), and (\ref{eq:hofmann}) signal the presence of entanglement.

\begin{figure}
\includegraphics[width=\columnwidth]{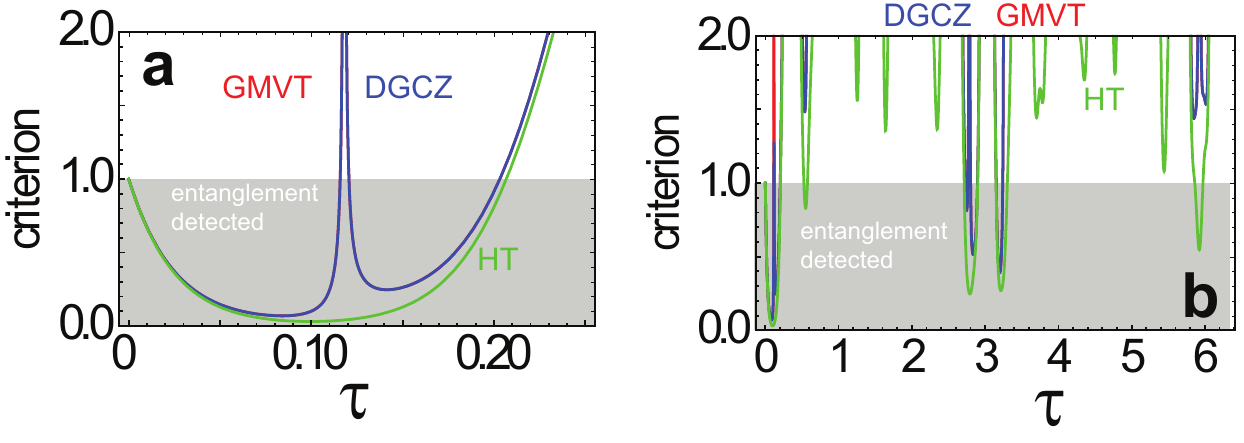}
\caption{Entanglement criteria for the two-axis two-spin squeezed state for a (a) short time range, (b) long time range.   Three entanglement criteria are calculated, the Giovannetti-Mancini-Vitali-Tombesi (GMVT) criterion (\ref{eq:Giovannetti}), the Duan-Giedke-Cirac-Zoller (DGCZ) criterion (\ref{eq:duan}), and the Hofmann-Takeuchi (HT) criterion  (\ref{eq:hofmann}).
Entanglement between the two BECs is detected for all values below the separability bound, represented by the shaded area. \label{fig7} }
\end{figure}

Figure \ref{fig7}(a) plots the left hand sides of the criteria (\ref{eq:Giovannetti}), (\ref{eq:duan}), and (\ref{eq:hofmann})  as a function of the interaction time $\tau $, in the short time range.
Initially, all three criteria detect entanglement successfully. However, the GMVT and DGCZ criteria fail at times near the optimal squeezing time $ \tau_{\text{opt}}^{(\text{sq})} $ due to the fact that at these times $ \langle S^z_j \rangle $ becomes small.  The HT criterion does not fail in these region since it does not rely on a comparison with $ \langle S^z_j \rangle $.  For longer times (Fig. \ref{fig7}(b)), all criteria are generally less successful at detecting entanglement, only detecting some specific time regions. For times which show strong squeezing, the entanglement criteria are able to detect entanglement.  The most successful criterion is the HT criterion, hence for this particular state (\ref{eq:hofmann}) appears to be the method of choice for correlation-based entanglement detection.     

\section{Wigner functions} 
\label{sec:wigner}

\subsection{Definitions}

In this section, we further analyze the 2A2S squeezed state by visualizing its Wigner function. The Wigner function represents the state as a quasiprobability distribution on the Bloch sphere and for two-mode BECs can be defined as \cite{dowling1994wigner}
\begin{align}
W(\theta , \phi ) = \sum _{l = 0}^{2j} \sum _{m = -l}^l \rho _{lm} Y_{lm}(\theta, \phi) ,
\label{Wigner}
\end{align}
where $ j = N/2 $, the spherical harmonics $Y_{lm}(\theta, \phi)$, and $\rho _{lm}$ given by
\begin{align}
 \rho_{lm} = \sum _{m_1 = -j}^j & \sum _{m_2 = -j}^j (-1)^{j-m_1 -m} \nonumber \\
& \times \left\langle j m_1 ; j -m_2 | lm \right\rangle \left\langle jm_1 | \rho | jm_2 \right\rangle  .
\end{align}
Here, $\rho$ is the density operator of the state being analyzed and $\left\langle j_1 m_1 ; j_2 m_2 | JM \right\rangle$ are the Clebsch-Gordan coefficients coupling two angular momentum eigenstates $\left| jm \right\rangle$. These states can be expressed in terms of the $S^z$ eigenstates using the relation
\begin{align}
\left| jm \right\rangle = \left| k = j+m \right\rangle  .
\end{align}
The Wigner function as in (\ref{Wigner}) is defined for a two-mode BEC with fixed atom number and thus cannot be directly applied to our case of two BECs involving four modes. Instead, we calculate the marginal and conditional Wigner functions where a tracing or projective operation is performed to obtain the Wigner functions on a single BEC.

\subsection{Marginal Wigner functions}

\begin{figure}
\includegraphics[width=\columnwidth]{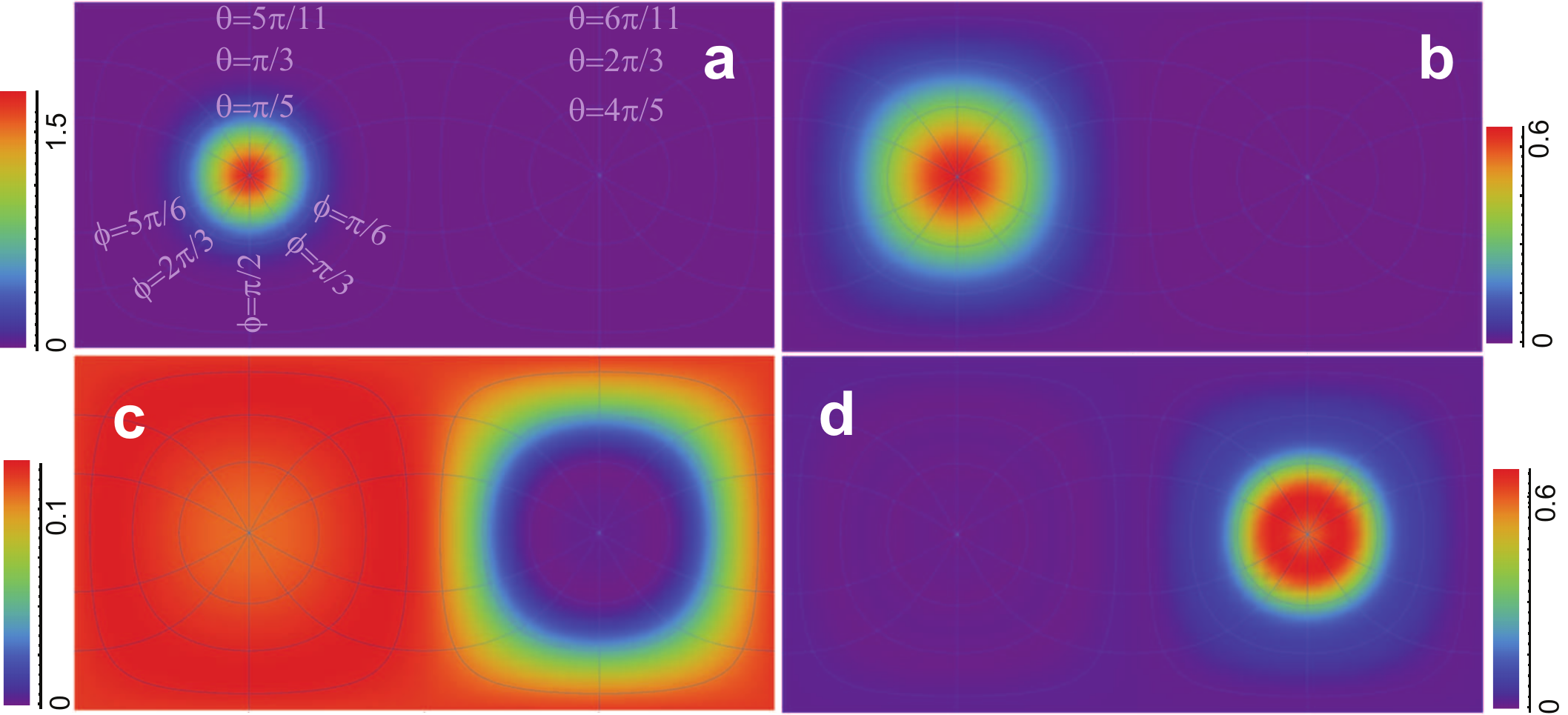}
\caption{Marginal Wigner function for four different interaction times (a) $\tau =0$;  (b) $\tau = \tau_{\text{opt}}^{(\text{sq})}/2 $;  (c) $\tau = \tau_{\text{opt}}^{(\text{sq})} $; (d) $\tau =2 \tau_{\text{opt}}^{(\text{sq})} $. 
The Bloch sphere is projected on a plane using the Cassini projection according to the transformation $ \theta = \sin^{-1} (\sin x \cos y) + \pi/2 $ and 
$ \phi = \tan^{-1} (\frac{\cos x}{\tan y}) $, where $ -\pi < x < \pi $ and $ -\pi/2 < y < \pi/2 $ are the coordinates on the Cassini projected map.     The number of atoms in each BEC is $N = 10 $ for all plots, where $ \tau_{\text{opt}}^{(\text{sq})} = 0.165 $. \label{fig8} }
\end{figure}

To calculate the marginal Wigner function, we take the partial trace of the two-BEC state (\ref{state}) over BEC 1 to obtain the reduced density matrix $ \rho_2 $.  
The state  (\ref{eq:rhotraced}) is then used to calculate the Wigner function according to (\ref{Wigner}).  Figure \ref{fig8} shows the marginal Wigner function for four different interaction times $\tau $. The initial state (Fig. \ref{fig8}(a)) starts as a Gaussian at the north pole of the Bloch sphere.  As the 2A2S Hamiltonian is turned on (Fig. \ref{fig8}(b)), the diameter of the Gaussian increases, until the distribution nearly covers the whole Bloch sphere at the optimal squeezing time $ \tau_{\text{opt}}^{(\text{sq})} $  (Fig. \ref{fig8}(c)).  At this point the average value of all spin variables is small, in agreement with Fig. \ref{fig2}.  For longer squeezing times, the probability distribution becomes more concentrated at the south pole of the Bloch sphere (Fig. \ref{fig8}(d)), at which point $ \langle S^z \rangle $ turns negative as seen in Fig. \ref{fig2}. The flipping between the north and south poles continues for longer times, as seen in Fig. \ref{fig2}(b). The Wigner function is completely rotationally symmetric around the $ S^z $ axis at all times.  In a similar way to the two-mode squeezed state, there is no squeezing for a single BEC, and the squeezing only appears in variables involving both BECs.

\subsection{Conditional Wigner functions}

For the conditional Wigner function, we first project the two-BEC state (\ref{state}) 
onto a Fock state (\ref{thetaphifock}) of one of the BECs (say, BEC 1), for a particular basis specified by $ (\theta, \phi) $.  This would correspond to performing a Fock state measurement in a given basis on BEC 1, where a random collapse occurs, then examining the measured state of BEC 2.  
The resulting quantum state is
\begin{align}
| \psi_{k} (\theta, \phi, \tau) \rangle = \frac{P_{k} (\theta, \phi) | \psi(\tau) \rangle }{\sqrt{\langle \psi(\tau) | P_{k} (\theta, \phi)  | \psi (\tau) \rangle }} ,
\label{projectedstate}
\end{align}
where the projector onto a Fock state on BEC 1 is
\begin{align}
P_{k} (\theta, \phi) = | k \rangle_1^{(\theta, \phi)} \langle k |_1^{(\theta, \phi)} .
\label{projectorcondwig}
\end{align}
To calculate the conditional Wigner functions, the state  $ \rho = | \psi_{k} (\theta, \phi, \tau) \rangle  \langle \psi_{k} (\theta, \phi, \tau) | $ is substituted into (\ref{Wigner}). 

\begin{figure}
\includegraphics[width=\columnwidth]{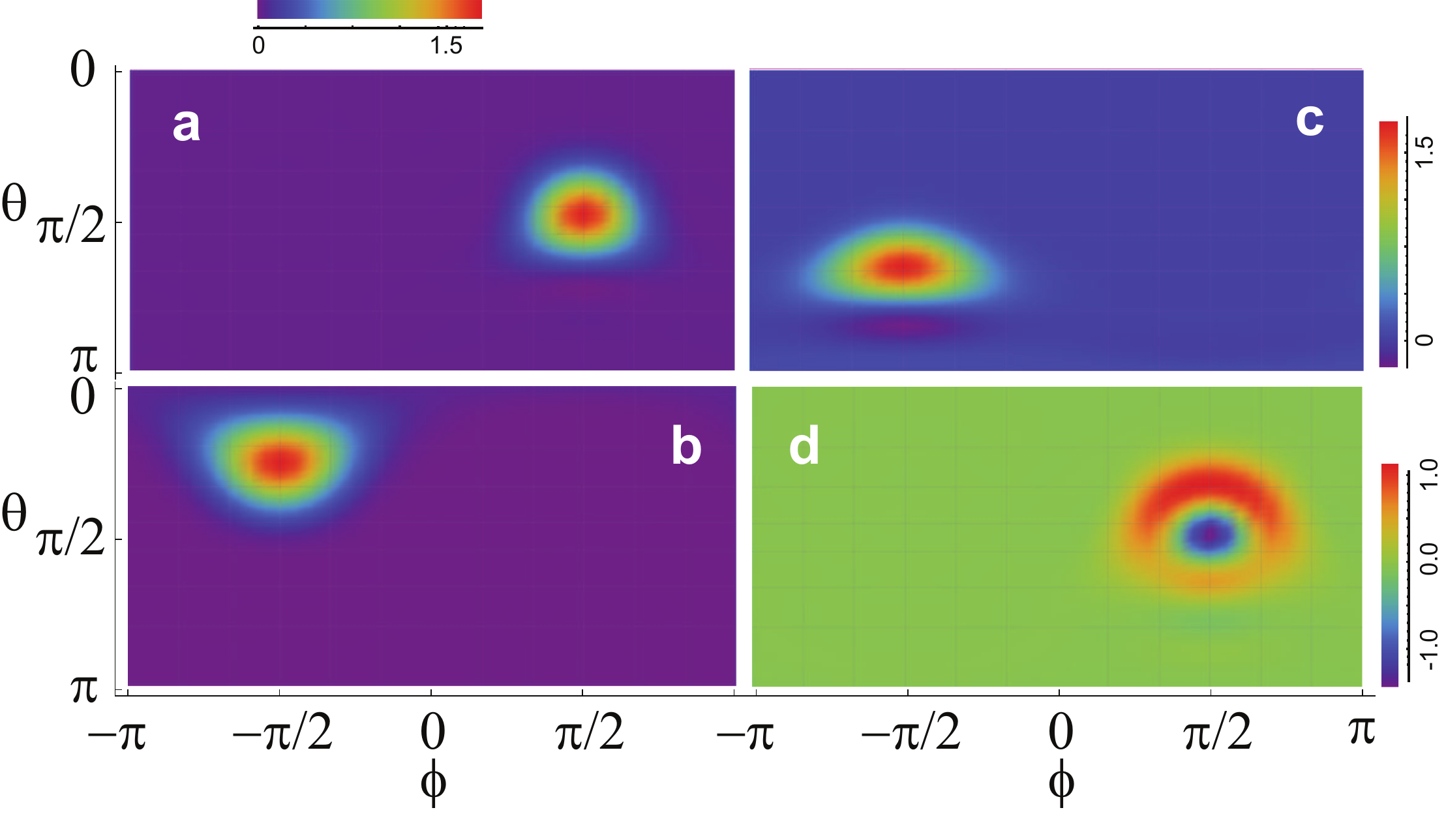}
\caption{Conditional Wigner function for after projecting the 2A2S squeezed state at the time $ \tau = \tau_{\text{opt}}^{(\text{sq})} $.  
The Wigner function is plotted with a Mercator projection in the basis $ (\tilde{S}^x, \tilde{S}^y, S^z ) $.  For example, the state such that 
$ \langle \tilde{S}^x \rangle = N $ has a Wigner function peaked at $ \theta = \pi/2, \phi = 0 $.  The projection parameters in (\ref{projectedstate}) are 
(a) $  \theta = \pi/2, \phi = \pi/2, k = N $; (b) $  \theta = \pi/4, \phi = -\pi/2, k = N $; (c) $  \theta = \pi/4, \phi = \pi/2, k = 0 $; (d) $  \theta = \pi/2, \phi = \pi/2, k = N-1 $.  The legends are shown adjacent to each subfigure except for (a) and (b), which are common. 
The number of atoms in each BEC is $N$ = 10 for all plots.
\label{fig9} }
\end{figure}

The resulting conditional Wigner functions for projections in various bases are shown in Fig. \ref{fig9}.  From the form of the spin-EPR state (\ref{prototypeepr}), we expect that the resulting state is 
\begin{align}
| \psi_{k} (\theta, \phi, \tau_{\text{opt}} ) \rangle \approx | k \rangle_2^{(\theta, \pi-\phi)}
\label{projectedstatecond}
\end{align}
on BEC 2.   For projections with $ k = N $ (Figs. \ref{fig9}(a)(b)), the projected state appears as a Gaussian Wigner function, characteristic of a spin coherent state, centered at the angular parameters  $ (\theta, \pi-\phi) $. This is in agreement with (\ref{projectedstatecond}), using the fact that  
\begin{align}
| k=N \rangle^{(\theta, \phi)} = | \theta, \phi \rangle \rangle .
\end{align}
Similarly, for a projection with $ k = 0 $  the resulting state is a spin coherent state centered at $ (\pi-\theta, \phi-\pi) $, using the fact that 
\begin{align}
| k=0 \rangle^{(\theta, \phi)} = | \pi - \theta, -\phi \rangle \rangle .
\end{align}
In the case shown in Fig. \ref{fig9}(c), the resulting spin coherent state is more distorted than those shown in Fig. \ref{fig9}(a)(b) because parameters are chosen such that the final state is located at $ \theta = 3\pi/4, \phi = -\pi/2 $. From the $ (S^z_1, S^z_2) $ plot in Fig. \ref{fig4}(b), we see that the probability distribution tends to diminish for negative $ S^z $, i.e. near the south pole of the Bloch sphere.  We attribute the distortion of the distribution in Fig. \ref{fig9}(c) to the generation of an imperfect spin-EPR state due to the relatively small ensemble sizes $ N = 10 $ considered here.

Finally, projecting on $ k = N-1 $ (Figs. \ref{fig9}(d)) produces a single particle Fock state centered around $ (\theta, \pi-\phi) $.  The resulting state shows a distribution that is similar to that of a single particle Fock state, with a negative central region, surrounded by a positive ring. Again, deviations from the exact spin-EPR state cause some differences to an ideal Fock state Wigner function.  For larger ensembles we expect that the distributions will more closely follow the distributions of the Fock states.

\section{Bell's inequality} 
\label{sec:bell}

In this section, we show that the 2A2S squeezed state violates a Bell inequality using only experimentally accessible correlations of total spin operators (\ref{eq:spinops}). Recently, the violation of a multipartite Bell inequality has been experimentally achieved with a single two-mode BEC \cite{schmied2016bell}. The measurement of violations of a bipartite Bell inequality using two spatially separated BECs, however, has not been reported yet. Previous studies on 1A2S entangled states suggested that parity measurements (i.e. $ \pm 1 $ depending on whether $ | k \rangle $ is even or odd) are required for violating a Bell inequality \cite{oudot2019bell}.  Parity measurements are currently experimentally challenging because they are sensitive at the single atom level. Methods to violate a Bell inequality without the use of parity measurement are therefore of interest.  

We use the Bell-CHSH inequality \cite{chsh1969} for two observers with two local measurement choices. Every local realist theory must satisfy 
\begin{align}
{\cal C} = \Big| & \langle M_1^{(1)} M_2^{(1)} \rangle + \langle M_1^{(1)} M_2^{(2)} \rangle \nonumber \\
&  - \langle M_1^{(2)} M_2^{(1)} \rangle + \langle M_1^{(2)} M_2^{(2)} \rangle \Big| \leq 2 ,\label{eq:CHSH}
\end{align}
where $M_1^{(i)}, M_2^{(j)} $ are local measurement choices on BEC 1 and BEC 2 respectively, and $i,j \in \{ 1, 2 \}$ label the two measurement choices. The measurement operators leading to a violation of (\ref{eq:CHSH}) are given by
\begin{align}
M_n^{(i)} = \sgn(\tilde{S}^{x}_n \cos \theta_n^{(i)} + \tilde{S}^{y}_n \sin \theta_n^{(i)} ) .
\label{belloperators}
\end{align}

Here the meaning of the $\sgn$-function is that the sign of the eigenvalues of the spin operators are taken:
\begin{align}
\sgn(\tilde{S}^{x} \cos \theta + \tilde{S}^{y} \sin \theta ) & = \sum_{k=0}^N \sgn(2 k - N) | k \rangle^{(\theta,0)} \langle k |^{(\theta,0)}  ,
\end{align}
where $ | k \rangle^{(\theta,\phi)} $ are the eigenstates of the rotated spin operator $ \tilde{S}^{x}_1 \cos \theta + \tilde{S}^{y}_1 \sin \theta  $ and is defined in (\ref{thetaphifock}).  Taking the $\sgn$-function of the rotated operators produces dichotomic (i.e. two-valued) measurement outcome as needed for the CHSH inequality. But importantly, this way of mapping the collective spin measurement to a dichotomic observable does not require single-atom precision, 
since the variation of the eigenvalues are split into two broad regions, $ k< N/2 $ and $ k > N/2 $.  

The choices of the parameters $ \theta_n^{(i)} $ and the interaction time $ \tau $ should generally be optimized such that a maximal violation is obtained. Firstly, we have found that the largest violations occur for interaction times  $ \tau \approx \tau_{\text{opt}}^{(\text{sq})} $. This is reasonable from the point of view that the Bell-CHSH inequality (\ref{eq:CHSH}) is correlation based, and the maximal squeezing tends to maximize the correlations.  For the angular parameters, we have found that a strong violation can be found using the parameter choices
\begin{align}
\theta_1^{(1)} & = 0 \nonumber \\
\theta_1^{(2)} & = \theta_B \nonumber \\
\theta_2^{(1)} & = \theta_B /2 \nonumber \\
\theta_2^{(2)} & = -\theta_B /2 .
\label{thetabdefs}
\end{align}
For the qubit case $ N = 1 $, the above basis choice reduces to the well-known optimal CHSH measurements by taking $ \theta_B = \pi/2 $. For larger values of $ N $, the sign-binned operators (\ref{belloperators}) no longer coincide with the spin operators. We find that the Tsirelson bound can no longer be reached for $N > 1$, possibly due to the loss of information caused by the binning process. We do, however, find violations of (\ref{eq:CHSH}) for angles $\theta_B$ significantly smaller than $\pi /2$. Both the violations and the optimal angle $\theta_B$ diminish with increasing $N$.

\begin{figure}
\includegraphics[width=\columnwidth]{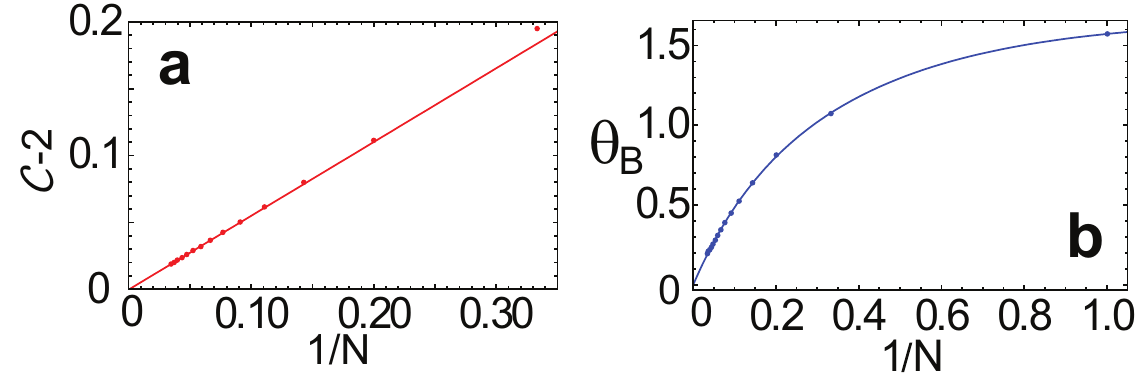}
\caption{(a) Optimized violation of the Bell-CHSH inequality $ {\cal C} - 2 $ as a function of $ 1/N $ for the 
2A2S squeezed state.  Dots show the numerically calculated values and the solid line is the linear fit (\ref{violationfit}).   (b) Optimal Bell angles $ \theta_B $ with the basis choices (\ref{thetabdefs}) as function of $ 1/N $.  Dots show the numerically calculated values and the solid line is the Pad{\'e} approximant (\ref{thetabpade}).  Interaction times of $ \tau = \tau_{\text{opt}}^{(\text{sq})} $ are used throughout. 
\label{fig10} }
\end{figure}

Figure \ref{fig10}(a) shows the optimized Bell-CHSH violations with respect to the angles $\theta_B $. The amount of violation is found to empirically have scaling that agrees very well with the relation 
\begin{align}
{\cal C} \approx 2 + \frac{0.55}{N} .
\label{violationfit}
\end{align}
The fact that the effect diminishes for large $ N $ is  expected as our system can be described by continuous variable mode operators in the limit of large $N$, 
according to the Holstein-Primakoff approximation.  For two-mode squeezed states, it is known that no Bell violation is possible with measurements that are linear combinations of quadrature operators \cite{braunstein2005quantum}. Maximal violations of the CHSH inequality with EPR states have been reported with parity measurements, e.g. in Ref. \cite{Chen2002EPRBell}, suggesting that single-atom resolution measurements may be needed to achieve maximal violations with the spin-EPR state.

In the limit of large $ N $ we therefore expect that the Bell-CHSH expression with the coarse-grained measurement operators (\ref{belloperators}) approaches 2, which can be attained if $ \theta_B = 0 $, such that all four correlators are $ \langle \sgn ( S^x_1  S^x_2) \rangle \rightarrow 1 $.  Interpolating small and large $ N $, and from the highly linear nature of Fig. \ref{fig10}(a), we expect that violations can be found for any finite value of $ N $, although they will become smaller and increasingly challenging to measure for larger $ N $.  

The optimal angles to violate the Bell-CHSH inequalities are given in Fig. \ref{fig10}(b). The angle does not follow a simple power law relation, and thus we fit it with a Pad{\'e} approximant
\begin{align}
\theta_B \approx \frac{6.1/N-0.67/N^2}{1+2.45/N},
\label{thetabpade}
\end{align}
which gives a close approximation to the numerically obtained angle.

\section{Summary and conclusions}
\label{sec:summary}

We have examined the 2A2S squeezed state from multiple points of view: squeezing, spin expectation values, probability distributions, entanglement, Wigner functions, and Bell correlations.  A consistent picture emerges from studying various quantities.  Starting from two BECs which are polarized in the positive $ S^z $-direction, the 2A2S Hamiltonian produces the spin-EPR state $ | \text{EPR}_{-} \rangle $, the analogue of the two-mode squeezed state for spins. This state then evolves to the state of two BECs polarized in the negative $ S^z $-direction, from which then the spin-EPR state $ | \text{EPR}_{+} \rangle $ is produced.  After returning to two BECs polarized in the positive $ S^z $-direction, the process repeats itself.  The picture that we describe here is not exact, even for the decoherence-free case that was examined in this paper. As can be seen in Figs. \ref{fig1}, \ref{fig2}, \ref{fig5}, and \ref{fig6} the oscillations are not perfectly periodic in amplitude or period. For such imperfect oscillations one may expect that the oscillations die out relatively quickly.   However, the oscillations are remarkably persistent for the times that we have examined, and the fidelities to the spin-EPR states remain high after a large number of oscillations.

The optimal times to produce the largest amounts of squeezing, entanglement, Bell correlations, and fidelities with the spin-EPR states are found to be similar, but not precisely the same.  Thus each quantity must be optimized separately in order to obtain the optimal values. Approximate formulas for the optimal times were obtained using a suitable fitting function which should be accurate particularly for large $N$  since it is determined by interpolating between numerically determined intermediate $N $ and $N \rightarrow \infty $.  It is found that optimal times for the entanglement and fidelities with spin-EPR states are closely approximated by the times that $ \langle S^z \rangle = 0 $.  Meanwhile the optimal times to violate the Bell-CHSH inequality are closest to the optimal squeezing time.   Since the measurement of $ \langle S^z \rangle $ is relatively simple, this is a convenient heuristic that can be used to generate the desired state.

One interesting feature of the 2A2S squeezed states is that they are able to violate a Bell-CHSH inequality, and it appears that this is possible for all finite $ N $. We deduce this from the fact that in the limit of large $ N $, the 2A2S squeezed state should approach a two-mode squeezed state.  The level of violation is rather small, dropping off as $ 1/N $, which makes it more difficult to observe for large ensembles.  However, we conjecture that this may be the best that can be done for a two measurement, two outcome CHSH inequality that only uses first order correlators of coarse-grained collective spin operators, since it is known that quadrature operator measurements alone cannot detect nonlocality in two-mode squeezed states \cite{braunstein2005quantum}.

\begin{acknowledgments}
This work is supported by the Shanghai Research Challenge Fund; New York University Global Seed Grants for Collaborative Research; National Natural Science Foundation of China (61571301,D1210036A); the NSFC Research Fund for International Young Scientists (11650110425,11850410426); NYU-ECNU Institute of Physics at NYU Shanghai; the Science and Technology Commission of Shanghai Municipality (17ZR1443600,19XD1423000); the China Science and Technology Exchange Center (NGA-16-001); and the NSFC-RFBR Collaborative grant (81811530112). 
\end{acknowledgments}

\appendix

\section{Expression for probability of two-axis two-spin state measured in various spin bases}
\label{app:prob}

The Fock states of the spin operators corresponding to $ \tilde{S}^x $ and $ \tilde{S}^y $ are given by
\begin{align}
| k  \rangle^{(\tilde{x})} & = e^{-iS^z \pi/8} e^{-iS^y \pi/4} | k \rangle^{(z)} \nonumber \\  
| k \rangle^{(\tilde{y})} & = e^{-iS^z 3\pi/8} e^{-iS^y \pi/4} | k \rangle^{(z)} .
\end{align}
Here the matrix elements of the $ S_y $ rotation are given by
\begin{align}
& \langle k | e^{-i S^y \theta/2} | k' \rangle = \sqrt{ k'! (N-k')! k! (N-k)!} \nonumber \\
& \times 
\sum_n \frac{(-1)^n \cos^{ k- k' + N - 2n} (\theta/2) \sin^{2n + k' - k} (\theta/2) }{(k-n)!(N-k'-n)!n!(k'-k+n)!} , 
\label{syrotmatrixelement}
\end{align}
where $ | k \rangle = | k \rangle^{(z)} $.

\section{Entanglement scaling under the Holstein-Primakoff approximation}
\label{app:hpentanglement}

In this section, we derive the entanglement at the optimal times under the Holstein-Primakoff approximation.  Starting from the Holstein-Primakoff approximated Hamiltonian (\ref{twomodesqueezeham}), the two-mode squeezed wavefunction is given as \cite{braunstein2005quantum}
\begin{align}
e^{-i H_2 t} | 0 \rangle  = c \sum_{n=0}^\infty  d^n |n\rangle |n \rangle
\label{generaltwomodesq}
\end{align}
where the coefficients are 
\begin{align}
c & = 1- \tanh^2 (N \tau ) \approx 4 e^{-2N\tau }\nonumber \\
d & = i \tanh^2 (N \tau) \approx i (1- 2e^{-2N\tau} )^2 .  
\end{align}
We take the time to be the right hand side of (\ref{hpvalidtime}). Evaluating the entropy for the state (\ref{generaltwomodesq}) gives a general expression
\begin{align}
E = \frac{c( |d| \log_2 (c/|d|)-\log_2 c) }{(|d|-1)^2} .  
\end{align}
Substituting the approximated values gives
\begin{align}
\frac{E}{E_\text{max}} & \approx \frac{4 ( (1-2N)^2 - N (1-4N) \ln N)}{(1-4N)^2 \ln (N+1)} \nonumber \\
& \approx 1 + \frac{1}{\ln N },
\end{align}
where in the second line we further approximated the expression for large $N $.  We see that for large $ N $, $ E/E_\text{max} $ logarithmically approaches 1.

\section{Basis invariance of the spin-EPR states}
\label{app:epr}

\subsection{Zero phase spin-EPR state}

First let us consider the spin-EPR state using the standard phase convention 
\begin{align}
|\text{EPR}_+ \rangle = \frac{1}{\sqrt{N+1}} \sum_{k=0}^N | k \rangle^{(\theta, \phi)} | k \rangle^{(\theta, -\phi)}  ,
\label{spineprgeneric}
\end{align}
where the rotated Fock states are defined in (\ref{thetaphifock}).  The state (\ref{spineprgeneric}) is not the same state (\ref{spinEPRxbasis}) as that is initially produced by the 2A2S Hamiltonian, we consider the above form first to simplify the presentation.  We will relate (\ref{spineprgeneric}) to the 2A2S case in the next section.  The aim here is to show that the spin-EPR states (\ref{spineprgeneric}) are independent of the choice of the basis angles $ \theta, \phi $. 

Consider the matrix element of the spin-EPR state
\begin{align}
& \left( \langle k | \otimes \langle k'| \right) | \text{EPR}_+ \rangle \nonumber \\
 & = \frac{1}{\sqrt{N+1}} \sum_{k'''=0}^N  \langle k |  e^{-i S^z \phi/2} e^{-i \tilde{S}^y \theta/2}  | k''' \rangle \nonumber \\
& \times \langle k' |  e^{i S^z \phi/2} e^{-i \tilde{S}^y \theta/2}  | k''' \rangle \nonumber \\
& =  \sum_{k'''=0}^N  \frac{e^{i (k'-k) \phi}}{\sqrt{N+1}}  \langle k |   e^{-i \tilde{S}^y \theta/2}  | k''' \rangle
\langle k' |  e^{-i \tilde{S}^y \theta/2}  | k''' \rangle ,
\end{align}
where we have evaluated the $ S^z $ phases to the left.  Now note that the matrix element $ \langle k |   e^{-i \tilde{S}^y \theta/2}  | k''' \rangle $ is real from (\ref{syrotmatrixelement}).  We may then write
\begin{align}
& \left( \langle k | \otimes \langle k'| \right) | \text{EPR}_+ \rangle \nonumber \\
& = \frac{1}{\sqrt{N+1}} \sum_{k'''=0}^N   e^{i (k'-k) \phi}  \langle k |   e^{-i \tilde{S}^y \theta/2}  | k''' \rangle \langle k''' |   e^{i \tilde{S}^y \theta/2}  | k' \rangle \nonumber \\
& = \frac{1}{\sqrt{N+1}} e^{i (k'-k) \phi} \langle k |   e^{-i \tilde{S}^y \theta/2} e^{i \tilde{S}^y \theta/2}  | k' \rangle \nonumber \\
& = \frac{1}{\sqrt{N+1}} e^{i (k'-k) \phi} \langle k  | k' \rangle \nonumber \\
& = \frac{1}{\sqrt{N+1}} e^{i (k'-k) \phi} \delta_{kk'} \nonumber \\
& = \frac{1}{\sqrt{N+1}} \delta_{kk'} .
\end{align}
This is the coefficient of spin-EPR state when $ \theta = \phi = 0 $:
\begin{align}
|\text{EPR}_+ \rangle = \frac{1}{\sqrt{N+1}} \sum_{k=0}^N | k \rangle^{(z)} | k \rangle^{(z)}  .
\label{standartepr}
\end{align}
Thus starting from a state with Fock states in an arbitrary basis, we have obtained the equivalent expression in the $ S^z $ basis.  It then follows that 
\begin{align}
|\text{EPR}_+ \rangle & = \frac{1}{\sqrt{N+1}} \sum_{k=0}^N | k \rangle^{(\theta, \phi)} | k \rangle^{(\theta, -\phi)} \nonumber  \\
& = \frac{1}{\sqrt{N+1}} \sum_{k=0}^N | k \rangle^{(z)} | k \rangle^{(z)}  \nonumber  \\
& = \frac{1}{\sqrt{N+1}} \sum_{k=0}^N | k \rangle^{(\theta', \phi')} | k \rangle^{(\theta', -\phi')},
\end{align}
where $ \theta', \phi' $ are another choice of parameters.  This means that the state (\ref{spineprgeneric}) can be written equivalently for an arbitrary basis choice $ \theta, \phi $ of the Fock states.

\subsection{Two-axis two-spin EPR state}

We now relate the state (\ref{spineprgeneric}) to the state generated by the 2A2S Hamiltonian.  Taking the specific case of $ \theta = \pi/2, \phi = 0 $ in (\ref{spineprgeneric}), we have
\begin{align}
|\text{EPR}_+ \rangle & = \frac{1}{\sqrt{N+1}} \sum_{k=0}^N | k \rangle^{(x)} | k \rangle^{(x)} .
\end{align}
Comparing this to (\ref{spinEPRxbasis}), we observe that the difference is that the second BEC's labels must be changed to $ k \rightarrow N-k $.  This can be achieved using the transformation
\begin{align}
e^{iS^z \pi/2} | k \rangle^{(x)} = i^N | N-k \rangle^{(x)} ,  
\end{align}
where we ignore irrelevant global phase factors. 
This suggests that the 2A2S-EPR state can be related to the spin-EPR state in the standard phase convention according to 
\begin{align}
|\text{EPR}_{-}  \rangle = e^{i S^z_2 \pi/2} |\text{EPR}_+ \rangle . 
\label{connectionepr} 
\end{align}

From (\ref{connectionepr}), and using $ \theta = \pi/2, \phi = \pi/2 $ in (\ref{spineprgeneric}), we obtain the form (\ref{spinEPRybasis}), where we used the fact that
\begin{align}
e^{iS^z \pi/2} | k \rangle^{(y)} = (-1)^k | N-k \rangle^{(y)} .  
\end{align}
Similarly, using (\ref{connectionepr}) and $ \theta = 0, \phi = 0 $ in (\ref{spineprgeneric}), we obtain (\ref{spinEPRzbasis}).  Finally, using the fact that 
\begin{align}
e^{iS^z \pi/2} | k \rangle^{(\theta, -\phi)} = | k \rangle^{(\theta, \pi-\phi)} ,
\end{align}
we obtain the most general form (\ref{prototypeepr}).

% How to do the references:
%% 1) First uncomment the below and compile
\bibliographystyle{apsrev}
\bibliography{refs}
%% 2) Copy the .bbl file to below and comment out the above two lines.
%\begin{thebibliography}{28}

\end{document}